\documentclass[%
 reprint,
 prl,
 floatfix,
 amsmath,amssymb,
 aps, twocolumn,
 longbibliography,
]{revtex4-2}

\usepackage{graphicx}
\usepackage{appendix}
\usepackage{xcolor}
\usepackage{hyperref}
\allowdisplaybreaks[2]

\begin{document}
\title{Cumulant expansion approach to the dynamics of interacting Mössbauer nuclei\\after strong impulsive excitation}

\author{Miriam \surname{Gerharz}}
\affiliation{Max-Planck-Institut f\"ur Kernphysik, Saupfercheckweg 1, 69117 Heidelberg, Germany}

\author{J\"org \surname{Evers}}
\affiliation{Max-Planck-Institut f\"ur Kernphysik, Saupfercheckweg 1, 69117 Heidelberg, Germany}

\date{\today}

\begin{abstract}
    Recent progress in accelerator-based x-ray sources brings higher excitation of ensembles of M\"ossbauer nuclei closer to experimental feasibility. Nevertheless, modeling the quantum dynamics of the interacting nuclear ensemble after an impulsive excitation is still an open challenge.
    Here, we employ an expansion in the inter-nuclear correlations to derive a  set of nonlinear equations which is capable of efficiently modeling large nuclear ensembles for arbitrary degrees of excitation. As a key signature for higher excitation,  we identify a non-linear time-evolution of the nuclear dipole phase, which can be tuned via the scattering geometry, and interferometrically be measured. 
    Our analysis further predicts finite-size effects in the nuclear dynamics of small ensembles as an interesting as-yet unexplored observable.
    Our results provide important  guidance for future experiments aiming at the non-linear excitation of nuclei. We further envision the exploration of finite size-effects in M\"ossbauer spectroscopy with highest spatial resolution, i.e., small sample volumes.  
\end{abstract}

\maketitle

Mössbauer nuclei have already proven to be a versatile platform for studying quantum optics and related concepts, even though typical experiments so far have been restricted to the  low-excitation limit with less than one signal photon on average due to source limitations~\cite{Rohlsberger2010,Rohlsberger2012,PhysRevLett.111.073601,PhysRevLett.114.207401,PhysRevLett.114.203601,Haber2016,Haber2017,PhysRevLett.123.250504,Heeg2021,doi:10.1126/sciadv.adn9825,PhysRevLett.77.3232,PhysRevLett.66.2037,yoshida_quantum_2021,adaptiveOptics,PhysRevA.65.023804,Vagizov2014,Vagizov2013,Sakshath2017,PhysRevLett.133.193401,nazeeri2025couplingnucleartransitionsurface,Yamashita2024,Lohse2026,10.1063/5.0249167,doi:10.7566/JPSJ.90.084705}. X-ray free-electron lasers~\cite{
emma2010first,barletta2010free,ishikawa2012compact,decking2020mhz,amann2012demonstration,inoue2019generation,liu2023cascaded,nam2021high,Liu04032025} provide access to new excitation regimes  which are expected to advance the field~\cite{moessbauer_story_dreams,yoshida_quantum_2021}, and first experiments have already demonstrated some of the novel possibilities~\cite{chumakov2018superradiance,Shvydko2023,scandium_new,single-shot}. Towards higher excitation, more than 900 signal photons after a single x-ray excitation have  been observed recently~\cite{single-shot}, and it has been theoretically proposed that focusing the x-ray beam could significantly enhance the fraction of excited nuclei in the ensemble~\cite{lentrodt2024towards}. Further source development such as  x-ray free-electron laser oscillators~\cite{PhysRevLett.100.244802,Margraf2023,Rauer2025} could even bring full inversion of an ensemble of Mössbauer nuclei within reach~\cite{adams_scientific_2019,lentrodt2024towards}. 

A corresponding theoretical framework is therefore required to match the pace of experimental advances.
From a theoretical point of view, the nuclear ensemble forms an interacting many-body system. As such, the problem is closely related to other many-body systems from a theoretical point of view. A particular challenge is that in general the coherent interaction between the nuclei and the incoherent loss channels may operate on comparable time scales, such that neither dominates. This renders some of the well-known approaches less suitable (see, e.g., the discussion in~\cite{Mink2023}). 
In the low-excitation limit, the resulting equations can analytically be solved, giving rise, e.g., to the well-known response function formalism~\cite{Kagan1979,Hannon1999} and  quantum optical models~\cite{Heeg2013,Lentrodt2020b,PhysRevA.102.033710}.
Towards higher excitation, different theoretical approaches have been pursued. Typical experiments excite the nuclei  impulsively, fast as compared to all other time scales in the system. This allows one to separate and analytically solve the nuclear excitation~\cite{lentrodt2024towards,lentrodt2024excitationnarrowxraytransitions}. By contrast, the subsequent decay dynamics of the interacting ensemble of nuclei is much harder to treat. 
It has been suggested to use a perturbative approach  to characterize the nuclear dynamics in leading order beyond the linear case~\cite{lukas}. This approach allows one to derive experimental signatures for the non-linear excitation, but it fails towards higher excitation. Assuming that the dynamics is restricted to the fully symmetric  superradiant subspace, the dynamics can also be evaluated~\cite{heeg2016inducingdetectingcollectivepopulation}. However, this approach fails to capture the incoherent single-particle dynamics which evolves the system out of the symmetric subspace, but is crucial in nuclei due to the non-radiative internal conversion channel~\cite{Hannon1999,ralf}. Recently, it has  been suggested to use a matrix-product state approach~\cite{SCHOLLWOCK201196} to model the  many-body dynamics \cite{xiangjin}. This approach can handle higher excitation and allows one to analyze photon correlations, but only a small number of nuclei can be modeled without further approximations and entanglement between the particles is only partially captured.

In the following, we pursue a different approach, which exploits that in typical bulk target materials the coupling between the nuclei is moderate.
This suggests a perturbative expansion in the nuclear coupling, rather than an expansion in the degree of excitation or a truncation of the Hilbert space. One approach is the {\it cumulant expansion}~\cite{kubo1962generalized}, which is well-known in modeling many-body dynamics in general~\cite{10.1063/1.5138937,PhysRevA.78.022102,PhysRevResearch.5.013091,Kirton_2018,Krmer2015,PhysRevA.104.023702} but so far has not been applied to model  the dynamics of M\"ossbauer nuclei.

Here, we solve the nuclear many-body system using a cumulant expansion approach. We derive effective many-body equations valid for arbitrary degrees of excitation, which can efficiently be solved even for larger ensembles since their number scales linearly with that of the nuclei. Assuming further a translational invariance in an extended ensemble under homogeneous excitation conditions, we derive an effective single-particle non-linear equation of motion which captures the many-body dynamics in leading order of the cumulant expansion. It allows one to model the many-body system very efficiently, independent of the number of nuclei. Solving this equation, we show that towards higher excitation, the couplings between the nuclei imply a characteristic non-linear evolution of the nuclear dipole phase, and put forward a method to interferometrically measure it which exploits the geometry-dependence of the inter-nuclear couplings. By comparing with the case without translational invariance, we further uncover interesting finite-size effects in smaller ensembles which could also be experimentally explored.

{\it Initial state after excitation.}
Because typical x-ray pulses are short (ps-fs) compared to the nuclear dynamics ($\sim$\,ns), we can consider the initial state after the x-ray excitation to be prepared according to the area theorem as $\langle \sigma^{ee}_l\rangle(0)=\sin^2(\mathcal{A}/2)$ and $\langle{\sigma_l}^-\rangle(0)=\sin(\mathcal{A}/2)\cos(\mathcal{A}/2)\exp(i\vec{k}_\mathrm{in}\vec{r}_l)$~\cite{allen-eberly}. Here $\mathcal{A}$ describes the degree of excitation in terms of pulse area, $\vec{k}_\mathrm{in}$ is the incident wave vector and $\vec{r}_l$ the position vector, $\langle \sigma^{ee}_l\rangle(0)$ the initial excited-state population and $\langle \sigma^{-}_l\rangle(0)$ the initial coherence of nucleus $l$. Note that the nuclei are embedded in a solid state sample, such that their positions are fixed. A phase gradient is imprinted onto the nuclear dipoles during excitation due to the lattice spacing $a_0$ being larger than the wavelength $\lambda_0$ ($a_0\approx 3.3 \lambda_0$). Therefore, in the relevant case of a partially-excited system, the initial phases of the nuclei are already locked by the excitation,  by contrast to fully inverted systems which may be sensitive to the choice of seeds for the initial phases \cite{Ostermann2019}. In the low-excitation regime, this allows for single-photon superradiance~\cite{Rohlsberger2010}. Below, we show find that under these conditions, the interaction between the nuclei induces a dominant non-linear phase evolution, while a superradiant enhancement remains small.

{\it Decay dynamics of the nuclear many-body system.} 
The decay dynamics  can be modeled using a master equation for an ensemble of $N$ interacting nuclei resonant to the incident light~\cite{Heeg2013, Lentrodt2020b,PhysRevA.95.033818,Ficek_Swain,Kiffner_Vacuum_Processes}
\begin{align}
 \dot \rho =& \frac{1}{i\hbar}\left[ H, \rho \right] + \mathcal{L}[\rho]\, ,\\
 H =& -\hbar \sum_{n,m=1}^N J_{mn} \: \sigma^+_n \, \sigma^-_m\,,\label{me-coh}\\
 \mathcal{L}[\rho] 
 =& \sum_{n,m=1}^N \frac{{\Gamma}_{mn}}{2} \left( 2\sigma^-_m \rho \sigma^+_n  - \sigma^+_n \sigma^-_m \rho - \rho \sigma^+_n \sigma^-_m  \right )\,, \label{me-inc}
\end{align}
where $ \Gamma_{mn} = \Gamma_{mn}^\mathrm{rad} + \delta_{mn}\Gamma^\mathrm{IC}$, which is well-tested against experimental data in the so far explored low-excitation regime~\cite{yoshida_quantum_2021,Lentrodt2020b}.  Here, $\mathcal{L}[\rho]$ describes the incoherent dynamics, where $\Gamma=\Gamma_{nn}$ is the single-particle total decay rate ($\Gamma=4.7\,$neV for the ${}^{57}$Fe M\"ossbauer transition at $14.4$~keV) comprising a radiative ($\Gamma_{nn}^\mathrm{rad}$) and a dominating single-particle non-radiative  ($\Gamma^\mathrm{IC}$)  contribution. The other elements $J_{nm}$ and $\Gamma_{nm}^\mathrm{rad}$  with $n\neq m$ describe the coherent and the incoherent coupling rates between two nuclei $n, m$ which are derived by tracing out the modes of the environmental radiation field.

We proceed by deriving equations of motion for the expectation values of relevant single-particle operators $\sigma_l^{-}$ and $\sigma_l^{ee} = \sigma_l^{+}\sigma_l^{-} $, where the former characterizes the coherence  and the latter the excited-state population of atom $l$. Here, $\sigma_l^{\pm}$ are Pauli raising and lowering spin operators acting on atom $l$. We obtain
\begin{align}
\frac{d}{dt}  \langle  \sigma^-_l \rangle =&
- \frac{ \Gamma_{ll}}{2} \left \langle \sigma^-_l  \right\rangle\nonumber \\
& - \sum_{\substack{n=1\\ n\neq l}}^N \mathcal{C}^*_{nl} \left( \left \langle \sigma^-_n  \right\rangle
- 2  \left \langle  \sigma^{ee}_l\sigma^-_n  \right\rangle \right) \,, \label{spec-1a} \\
\frac{d}{dt}  \langle \sigma^{ee}_l \rangle =&-  \Gamma_{ll} \langle \sigma^{ee}_l \rangle \nonumber\\
&-\sum_{\substack{n=1\\n\neq l}}^N \mathcal{C}_{nl}  
\left\langle\sigma^+_n  \sigma^-_l \right\rangle 
+ \mathcal{C}^*_{ln} \left \langle \sigma^+_l \sigma^-_n  \right\rangle\,, \label{spec-1b}
\end{align}
where $\mathcal{C}_{mn} = \frac{{\Gamma}_{mn}}{2} + i J_{mn}$.

{\it Cumulant expansion.} 
The equations of motion for the single-particle expectation values Eqs.~(\ref{spec-1a}) and (\ref{spec-1b}) depend on two-particle expectation values. This hierarchy  generalizes to higher-order expectation values, rendering an exact solution impossible. In the cumulant expansion approach, a closed set of equations is obtained by truncating this hierarchy at a given order~\cite{kubo1962generalized}. We employ the first order expansion, by approximating
\begin{align}
\langle A_n\, B_m\rangle \approx \langle A_n\rangle \, \langle B_m\rangle\qquad (n\neq m)\,,
\end{align}
for operators acting on different atoms $n\neq m$. In general, the results may depend on the truncation order~\cite{PhysRevA.100.041602,Ostermann2019,PRXQuantum.5.010344}. However, because the couplings are comparably weak, and because the nuclear dipoles are already synchronized during excitation in contrast to studies on phase synchronization, e.g. in~\cite{Zhu_2015}, it is expected that the first order in our case already captures the relevant dynamics. A comparison with the second order is provided in the Appendix, which confirms that the relevant dynamics is indeed already captured by the first order calculation, and additional benchmarking with the continuous-discrete truncated Wigner approximation \cite{Mink2023} is provided in the Appendix. In first order, we obtain
\begin{align}
    \frac{d}{dt}  \langle  \bar \sigma^-_l \rangle 
&=- \frac{ \Gamma_{ll}}{2} \left \langle \bar \sigma^-_l  \right\rangle
-\left(1-2  \left \langle  \sigma^{ee}_l  \right\rangle \right)  \kappa_l^R\,, \label{cum-1-coh}\\
\frac{d}{dt} \phi_l &= -\left(1-2  \left \langle  \sigma^{ee}_l  \right\rangle \right)  \langle  \bar \sigma^-_l \rangle^{-1} \kappa_l^I\,, \label{cum-1-phase} \\
\frac{d}{dt}  \langle \sigma^{ee}_l \rangle
&= -  \Gamma_{ll} \langle \sigma^{ee}_l \rangle 
-2\, \left \langle \bar \sigma^-_l \right\rangle\, \kappa_l^R\,,
\label{cum-1-pop}\\
\kappa_l&= \kappa_l^R + i \kappa_l^I = \sum_{\substack{n=1\\ n\neq l}}^N \mathcal{C}^*_{nl}   \langle\bar \sigma^-_n\rangle e^{i(\phi_n - \phi_l)}\,.
\end{align}
Here, we have further decomposed the nuclear coherences into their magnitude and their phase,
 $\langle  \sigma^-_l \rangle =  |\langle   \sigma^-_l \rangle|\, e^{i\phi_l} =  \langle  \bar \sigma^-_l \rangle\, e^{i\phi_l}$.
We find from Eq.~(\ref{cum-1-phase}) that the phase $\phi_l$ of the nuclear dipole moments in general has a non-linear and excitation-dependent time evolution. Below we develop this feature as a signature for non-linear nuclear excitation.

Note that the Eqs.~(\ref{cum-1-coh})-(\ref{cum-1-pop}) already admit for an  efficient numerical simulation, since they have a favorable  linear scaling with the number of nuclei $N$. In contrast, the number of matrix elements of the original density operator scales exponentially with $N$. 

{\it Translationally invariant systems.} 
We now continue the discussion by focusing on translationally invariant systems, which are well-established for Mössbauer samples in the low excitation regime~\cite{Heeg2013,Lentrodt2020b}. This will allow us to reduce the computational effort further. To this end, we assume that the nuclei are arranged in an infinite translationally invariant lattice. Below, we will focus on one dimensional chains, as a simple model, e.g.,  for quasi-1D  iron nanowires~\cite{Grinter2023,Mohaddes-Ardabili2004,824421,10.1063/1.373001,PhysRevB.72.024428,JBWang_2004} or other approximately one-dimensional samples \cite{Erb2022,PhysRevLett.118.237204,Lohse2026}. In the Appendix, we also generalize our analysis to the two-dimensional case and find similar results. Furthermore, we assume that the initial excitation is due to a plane wave field incident with wave vector $\vec{k}$ on all nuclei with the same magnitude. Hence, the coherences of  all nuclei after the excitation differ only in their initial phases $\phi_l(0)$.  In this case, Eqs.~(\ref{cum-1-coh})-(\ref{cum-1-pop}) become
\begin{align}
    \frac{d}{dt}  \langle  \bar \sigma^- \rangle 
&=- \frac{ \Gamma}{2} \left \langle \bar \sigma^-  \right\rangle
-\left(1-2  \left \langle  \sigma^{ee}  \right\rangle \right) \langle  \bar \sigma^- \rangle K^R\,,\label{eq:coh_infinite}\\
\frac{d}{dt} \phi &= -\left(1-2  \left \langle  \sigma^{ee}  \right\rangle \right)  K^I\,, \label{eq:phase}\\
\frac{d}{dt}  \langle \sigma^{ee}\rangle
&= -  \Gamma \langle \sigma^{ee} \rangle 
-2\, \left \langle \bar \sigma^- \right\rangle\, K^R\,,
\label{cum-1-pop-trans}\\
K &   =K_l =  \sum_{\substack{n=1\\ n\neq l}}^N \mathcal{C}^*_{nl}    e^{i(\phi_n(0) - \phi_l(0))} \label{eq:K_finite}\,.
\end{align}
Here,  $\langle  \bar \sigma^- \rangle =  \langle  \bar \sigma^-_l \rangle$, $\langle  \bar \sigma^{ee} \rangle =  \langle  \bar \sigma^{ee}_l \rangle$ and $K=K_l$ are the same for all atoms, respectively, and $K^R$ and $K^I$ denote the real and imaginary part of $K$, respectively. Analogously, the nuclear phase evolves as $\phi_l = \phi_l(0) + \phi$ including the initial excitation phase. Finally, $K$ is a constant which crucially determines the dynamics.

As a result we find that for typical experimental situations, the many-body dynamics  can be modeled using a simple set of nonlinear equations, independent of the number of nuclei.
A conceptually similar study has been performed in \cite{Krämer_2016}. In this work, the authors also reduce the system to three equations with effective couplings however with the goal to minimize the collective energy shift, opposite of our maximizing of $K^I$.
In the Appendix, a comparison to  chains of finite length is provided, which supports the validity of our translationally invariant model.

\begin{figure}
    \centering
    \includegraphics[width=0.99\linewidth]{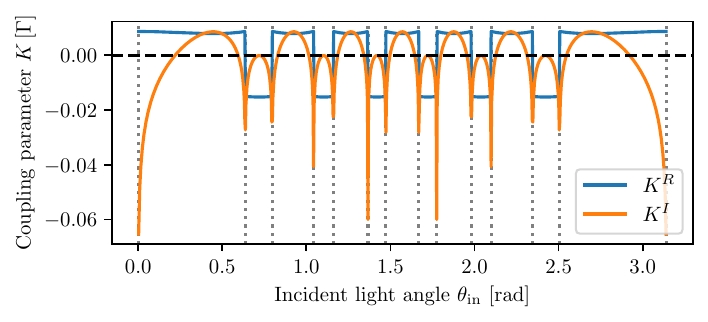}
    \caption{Coupling parameter $K=K^R+i K^I$ as function of the incident angle $\theta_\mathrm{in}$. For the calculation, Eq.~(\ref{eq:K_infiniteLinearChain}) is used with dipole orientation $\theta_d=\pi/2$. 
    }
    \label{fig:K_infiniteChain}
\end{figure}

{\it Temporal dynamics in the low-excitation limit.} 
Eqs.~(\ref{eq:coh_infinite})-(\ref{cum-1-pop-trans}) immediately allow us to connect our results to known phenomena in the well-studied low-excitation limit $\rho^{ee}\approx 0$. Then, the equations of motion reduce to
$\frac{d}{dt}  \langle  \bar \sigma^- \rangle 
\approx -  (\Gamma/2  + K^R)  \langle \bar \sigma^-  \rangle$
and
$\frac{d}{dt} \phi \approx -  K^I$.
%
%
From these equations we can immediately interpret the real and imaginary parts of $K$ in terms of a superradiant broadening and an interaction-induced  energy shift of the nuclear resonance, respectively~\cite{Rohlsberger2010,yoshida_quantum_2021,Hannon1999}.

{\it Temporal dynamics beyond the low-excitation limit.} 
Next we turn to the largely unexplored regime of higher excitation. We start by noting
that the magnitude of the coupling parameter $K$ generally is small as compared to $\Gamma$. The reason is that the total decay rate $\Gamma$ is dominated by the single-particle non-radiative (internal conversion) parts $\Gamma^{\mathrm{IC}}$ which do not contribute to the couplings (a more detailed numerical study of $K$ is provided below). 
Therefore  the phase of the nuclear dipole moment Eq.~(\ref{eq:coh_infinite}) is the most promising signature for effects of the non-linear excitation, as it is only governed by $K$, but not by $\Gamma$, in contrast to studies focusing on superradiance~\cite{zs9x-9x6f,PhysRevResearch.5.013091}. For this reason we will consider the phase of the nuclear dipole moment as the main observable throughout the manuscript. In order to derive its time evolution, we can approximately solve for  $\langle \sigma^{ee}\rangle$  by neglecting the small contribution $K^R$ in Eq.~(\ref{cum-1-pop-trans}). Inserting the solution $\langle\sigma^{ee}\rangle(t)\approx \sin^2(\mathcal{A}/2)\exp(-\Gamma t)$ into Eq.~(\ref{eq:coh_infinite}), we find 
\begin{equation} \label{eq:integratedPhase}
    \phi(t)=-K^I\left[ t - \frac{2}{\Gamma}\left(1-e^{-\Gamma t}\right)\sin^2\left(\frac{\mathcal{A}}{2}\right) \right] + \phi_0\,. 
\end{equation}
Here, $\mathcal{A}$ is the pulse area characterizing the initial  nuclear excitation via  $\langle \sigma^{ee}\rangle(0)=\sin^2(\mathcal{A}/2)$ and $\langle\bar{\sigma}^-\rangle(0)=\sin(\mathcal{A}/2)\cos(\mathcal{A}/2)$~\cite{allen-eberly}. 
We find that the nuclear phase evolves non-linearly at intermediate times  until $\langle \sigma^{ee}\rangle$ has decayed to zero, with magnitude governed by $K^I$. 

{\it Coupling parameter $K$}. In order to explore the range of the central quantity governing the coupled decay $K$, we assume that all nuclear dipole moments are aligned at a fixed angle $\theta_d$ relative to the chain axis, and that the incident plane wave x-ray pulse propagates at an angle $\theta_\mathrm{in}$ with respect to the chain. X-rays from an XFEL are naturally linearly polarized with good purity. We model the couplings using  free-space dipole-dipole couplings in vacuum~\cite{Ficek_Swain,Kiffner_Vacuum_Processes}, since the iron refractive index at the relevant x-ray energy 14.4~keV is close to unity (for more details see Appendix). $K$ then becomes (see Appendix)
\begin{align}
  K &=  -\frac{3i\Gamma^\mathrm{rad}}{2 \eta_0} \textcolor{black}{\sin^2(\theta_d)} \:\mathcal{X}_1  
\textcolor{black}{-}\frac{3\Gamma^\mathrm{rad}}{4\eta_0^2}[1+3\cos(2\theta_d)] \:\mathcal{X}_2 \nonumber \\
&\quad \textcolor{black}{-}\frac{3i\Gamma^\mathrm{rad}}{4\eta_0^3}[1+3\cos(2\theta_d)]\: \mathcal{X}_3 \,, \label{eq:K_infiniteLinearChain}
\end{align}
where $\mathcal{X}_n = \mathrm{Li}_n( \exp[i(\eta_0 +\Delta \phi)]) + \mathrm{Li}_n( \exp[i(\eta_0 -\Delta \phi)])$ with the polylogarithm $\mathrm{Li}_n(z)$. 
For the analysis, we use the scaled distance parameter $\eta_0=k_0 a_0\approx 21$ with wavevector $k_0=2\pi/\lambda_0$  and parameters for ${}^{57}$Fe ($\lambda_0 = 86$~pm) with nearest-neighbor spacing as in $\alpha$-Fe ($a_0 = 287$~pm). Further $\Delta \phi = \eta_0 \cos \theta_\mathrm{in}$ characterizes the incident phase difference between two neighboring nuclei. 

Results are shown in Fig.~\ref{fig:K_infiniteChain} as function of the x-ray incidence angle. We use an alignment of the dipole moments  $\theta_d=\pi/2$ because in this case the dominant $1/\eta$-term in $K$ is maximized.
Overall, we find that $K$ is comparably small on the scale of the total decay rate $\Gamma$, confirming our approximation leading to Eq.~(\ref{eq:integratedPhase}). 
For $K^I$, in the center around the perpendicular incident direction, a set of characteristic minima appears. Mathematically these can be traced back to formal divergencies of the polylogarithm for $\theta_d\neq0$ and $\eta_0 \pm \Delta \phi=n 2\pi$ with $n$ an integer. In practice, this divergence is regularized by a finite interaction volume, e.g., limited by the photo-absorption in the target. From an experimental point of view, the region of small incidence angles appears most favorable, since it enables tuning of $K^I$ over its entire range.  Small incidence angles also allow for a large number of nuclei in the excitation volume, as required by assumption of translational invariance. For example, a beam of  width $100\,\mu$m incident on  the chain under an incident light angle of $5$\,mrad couples to approximately 100 million nuclei.

\begin{figure}
    \centering
    \includegraphics[width=\linewidth]{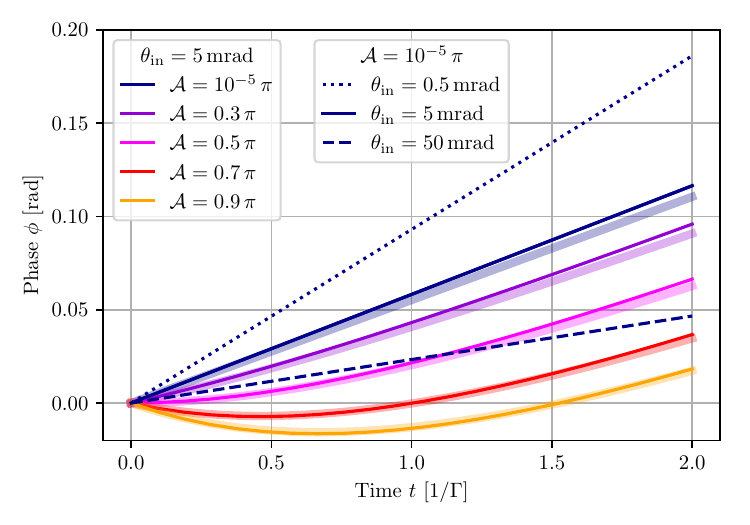}
    \caption{Nuclear dipole phase evolution as function of incidence angle and degree of excitation.  For a fixed incident angle $\theta_\mathrm{in}=5\,$mrad the different colors of the solid lines indicate different degrees of initial excitation $\mathcal{A}$. The transition to a non-linear phase evolution is clearly visible. For comparison, the shaded lines show the results for the central nucleus calculated on a finite chain with \textcolor{black}{$N=3000$} nuclei. In addition, in the low-excitation regime $\mathcal{A}=10^{-5}\pi$, the different line styles of the blue lines show the results for different incident light angles $\theta_\mathrm{in}$. The dipole moment is set perpendicular to the chain of nuclei ($\theta_d=\pi/2$).}
    \label{fig:phase}
\end{figure}

{\it Numerical simulation of the phase evolution.} Finally we  turn to a numerical simulation of the phase evolution governed by Eqs.~(\ref{eq:coh_infinite})-(\ref{cum-1-pop-trans}). The result is shown in Fig.~\ref{fig:phase}  for different degrees of initial excitations $\mathcal{A}$ and x-ray incidence  angles $\theta_\mathrm{in}$. 
The blue curves correspond to the limit of linear excitation ($\mathcal{A}=10^{-5}\pi$), with a  linear time evolution as expected. From the dotted and the dashed lines we further find that the slope of the phase can be controlled via the incidence angle $\theta_\mathrm{in}$, as described by $K^I$.
Next, we turn to the main results  beyond the linear excitation regime. The solid lines of different color show the evolution for different initial degrees of excitation $\mathcal A$ at x-ray incidence angle $\theta_\mathrm{in}=5$\,mrad. With increasing excitation, a non-linear time evolution of the phase develops, which leads to a growing deviation from the low-excitation case. At later times, the phase evolution becomes linear again, but  an excitation-dependent offset to the linear case remains. 

In order to quantify the impact of the assumption of an infinite chain of atoms, the shaded colored lines display the phase evolution of the central atom in a finite chain of $N=3000$ nuclei. The results qualitatively agree well with the infinite chain results. As discussed in the Appendix, the small quantitative differences can be  attributed to the deviation of the finite-chain coupling parameter $K$ to that of the infinite chain, which provide a handle to experimentally explore finite-size effects in nuclear ensembles. In the Appendix, we further compare our results to calculations based on the continuous-discrete truncated Wigner approximation~\cite{Mink2023}, an alternative approximation which is expected to work best in the case of strong nuclear correlations. We find good qualitative agreement, which supports the validity of the cluster expansion approach for the considered settings.

{\it Experimental signatures of the non-linear phase evolution.} 
Finally, we discuss how the non-linear phase evolution shown in Fig.~\ref{fig:phase} can be experimentally explored. We focus on the  scattered electric field, which is commonly observed in experiments with M\"ossbauer nuclei. The overall electric field is obtained from a suitable summation of the contributions of each of the nuclear dipole moments $\langle \sigma^-_l\rangle$. We consider a scattering geometry in which the differences between the initial phases $\phi_l(0)$ are compensated for by the respective propagational phases from the individual nuclei to the detector, e.g., the forward scattering geometry. In this case, the residual relevant phase evolution of all nuclei is given by $\phi(t)$.

We propose to measure the time evolution of the phase $\phi(t)$  interferometrically, which is well-established in the context of Mössbauer nuclei~\cite{Heeg2021,adaptiveOptics}, using two chains of nuclei.
The orientation of the first chain relative to the x-ray incidence direction  is varied, leading to phase evolutions as shown in Fig.~\ref{fig:phase}. The other reference chain is fixed at an incidence angle where $K^I\approx0$ ($\theta_\mathrm{in}\approx 0.22\,$rad). In this setting, the intensity of the x-ray scattered by both chains increases together with the degree of excitation, but only the first one exhibits a non-linear phase evolution, which is ideal for interferometry with high visibility.  
The effect of the different phase evolutions can then most straightforwardly be measured by detuning the resonance energy of the reference sample by an energy shift $\Delta$, e.g., via a M\"ossbauer drive~\cite{greenwood2012mossbauer}. The total intensity then takes the form
   $ I_\mathrm{comb}(t) \propto |A_\mathrm{ref}(t)e^{i\Delta \cdot t} + A_\mathrm{sample}e^{i\phi_\mathrm{sample}(t)}|^2$,
where $A_\mathrm{ref}$ and $A_\mathrm{sample}$ are the amplitudes of the scattered fields of the reference chain and the sample chain, respectively. Assuming that both amplitudes approximately follow the single-particle exponential decay $A_\mathrm{ref}(t)\approx A_\mathrm{sample}(t)\approx \exp(-\Gamma t/2)$, the total intensity approximately becomes
    $I_\mathrm{comb}\propto e^{-\Gamma t}\sin [\Delta \cdot t + \phi_\mathrm{sample}(t) ]$.
Hence,  the interference between the chains leads to a modulation of the exponentially decaying intensity with frequency $\Delta$. The non-linear phase shift of the first chain  shifts this entire quantum beat pattern, facilitating its  measurement.

\begin{figure}
    \centering
    \includegraphics[width=\linewidth]{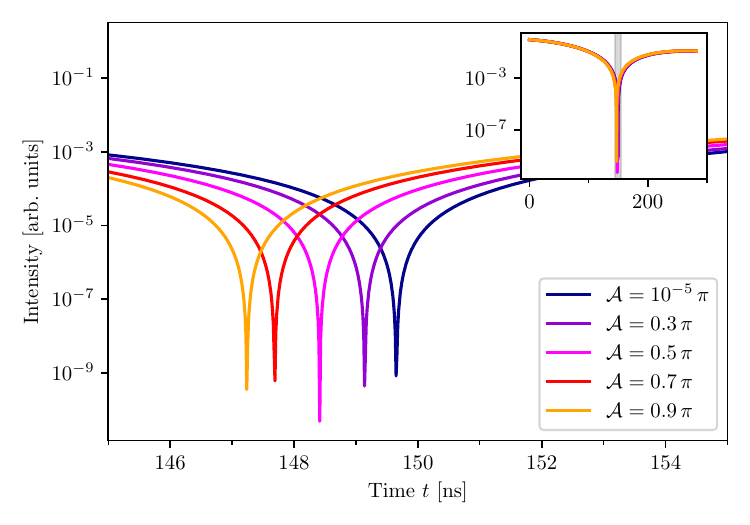}
    \caption{Time-dependent intensity of the x-ray scattered by two nuclear chains. The reference chain has a detuning of the nuclear resonance of $\Delta=-3\Gamma$, such that the interference between the chains leads to pronounced minima in the intensity. The non-linear phase evolution can then be observed via shifts of these minima in time.   The different lines illustrate the dependence on the different degrees of excitation $\mathcal{A}$. For the sample chain the dipole moment is set perpendicular to the chain of nuclei ($\theta_d=\pi/2$) and the x-ray incidence angles for the sample [reference] chain is $5\,$mrad [0.22\, rad]. The inset shows the signal intensity as a function of time over a larger time interval. The gray area marks the time region shown in the main panel.}
    \label{fig:interference}
\end{figure}

An example for the time-dependent intensity in this detection scheme is shown in Fig.~\ref{fig:interference}. The detuning is chosen as $\Delta=-3\,\Gamma$. The linear low-excitation case is given by the dark blue line with lowest value for $\mathcal A$. With increasing degree of excitation, the quantum beat minimum progressively shifts towards lower times by a few nanoseconds. Note that state-of-the-art experiments on nuclear forward scattering routinely measure the time-dependence with (sub-)nanosecond resolution. Alternatively, the nuclear phase variation could also be observed via the spectrum of the scattered light, in which already static phase shifts in the low-excitation regime are reflected in the observed line shape~\cite{PhysRevLett.114.207401}.

{\it Discussion.} Motivated by recent experimental progress in M\"ossbauer science towards stronger excitation, we analyzed the decay dynamics of an interacting ensemble of M\"ossbauer nuclei after an impulsive  excitation by an accelerator-based x-ray source. Unlike previous approaches, we  expanded the equations of motion to leading order in the degree of interaction-induced nuclear correlations, which is justified by the comparably small coupling of nuclei in bulk material. Within this approach, we  derived equations of motions for the ensemble which scale only linearly in the number of nuclei, and therefore can efficiently be simulated even for larger ensembles. Assuming further a translational invariance of the arrangement of the nuclei and the x-ray excitation, which is a good approximation for established experimental settings, we could further reduce the theoretical description to three non-linear  real-valued equations which describe the dynamics of the entire ensemble. We  found that the dynamics then is governed by a single coupling parameter $K$, which can be tuned in experiments.

As the main signature for higher excitation, we found that the inter-nuclear interactions lead to  additional non-linear phase dynamics of the nuclear dipole moments. Based on this observation, we proposed a setup for measuring  this signature, exploiting the geometry-dependence of $K$.  This opens up an approach for the exploration of  nuclear decay dynamics beyond the as-yet unexplored linear excitation regime, ideally with an analysis on the level of individual x-ray excitations~\cite{single-shot} to accommodate for source intensity fluctuations. 

For the future, we envision the generalization of our results to structured nuclear environments, which allow one to engineer and enhance the inter-nuclear couplings~\cite{yoshida_quantum_2021,oliver,PhysRevA.106.053701,Longo2016,Lohse2026}. Such design capabilities could allow one to achieve particular control operations for the nuclear phase, or to also significantly affect the nuclear population dynamics. 
Furthermore, source development also brings experiments on smaller samples within reach. We identified characteristic finite-size effects which already appear at low-excitation conditions, which are accessible with state-of-the-art XFELs. They lead to both, variations in the coupling parameter $K$ with chain length, as well as a dependence of the nuclear dynamics on their position within the chain. We envision future experiments exploring these finite-size effects, e.g., exploring the impact of the geometry on the coherence volume responsible for the collective effects observed in nuclear resonance scattering. Such studies would also form an important basis for future M\"ossbauer experiments with highest spatial resolution, i.e., small sample volumes.  


\begin{acknowledgments}
The authors would like to thank D. Lentrodt  for valuable discussions.
\end{acknowledgments}

\appendix
\section{Appendix}

\begin{figure}
    \centering
    \includegraphics[width=\linewidth]{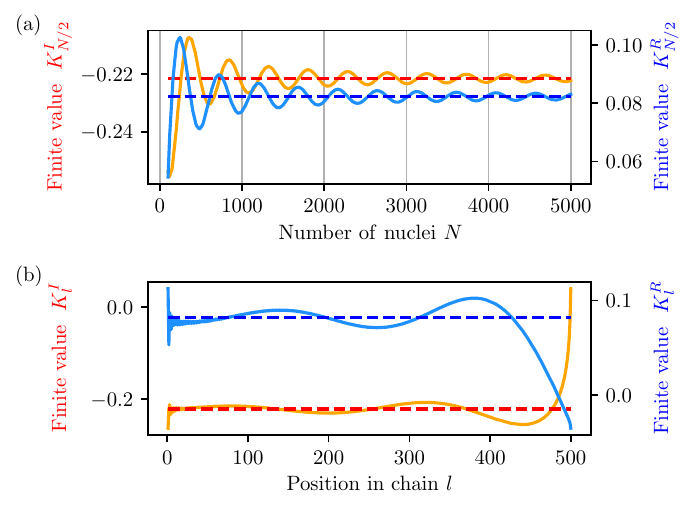}
    \caption{Comparison of the coupling parameter $K_l$ being evaluated on the finite chain according to Eq.~(\ref{eq:K_finite}) (solid) with that of the infinite chain according to Eq.~(\ref{eq:K_infiniteLinearChain}) (dashed). In (a) the coupling parameter $K$ evaluated at the central nucleus $l=N/2$ is shown as a function of the number of nuclei $N$. In (b) for a fixed number of nuclei $N=500$ the coupling parameter is displayed as function of the nucleus' position in the chain $l$. The calculations are performed for dipole moment perpendicular to the chain ($\theta_d=\pi/2$) and a small incident light angle of $\theta_\mathrm{in}=50$\,mrad.}
    \label{fig:convergenceOfK}
\end{figure}

\begin{figure}
    \centering
    \includegraphics[width=\linewidth]{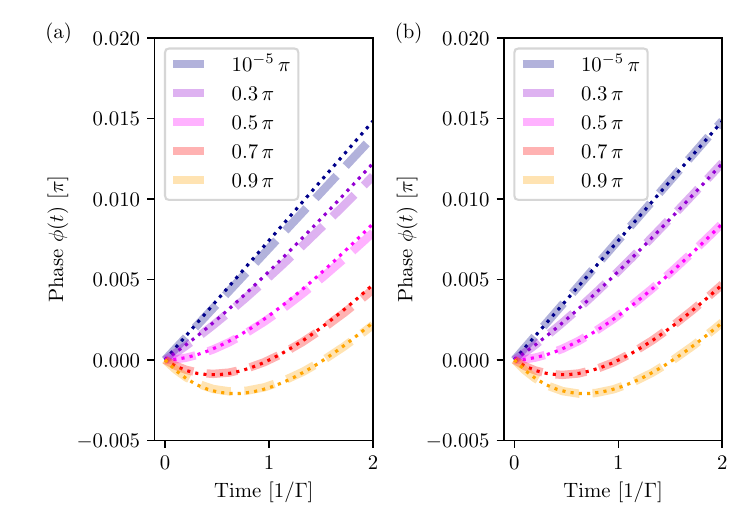}
    \caption{Comparison of the phase evolution computed with a finite chain evaluated at the central nucleus (dashed) and the translational invariant model (\textcolor{black}{dotted}). In (a) the chain has a length of $N=362$ nuclei, which corresponds to a point of maximal deviation of $K$ in Fig.~\ref{fig:convergenceOfK}, in (b) a length of $N=257$ nuclei corresponding to a point of minimal deviation in $K$ is chosen. The different colors indicate different initial excitations $\mathcal{A}$ according to the respective legend. The calculations are performed for dipole moment perpendicular to the chain ($\theta_d=\pi/2$) and a small incident light angle of $\theta_\mathrm{in}=50$\,mrad. }
    \label{fig:phase_finiteSizeEffects}
\end{figure}

\textit{Finite size effects.} In the main text we focus on larger translationally invariant systems, for which we evaluate the coupling parameter $K$ in the limit of an infinite chain Eq.~(\ref{eq:K_infiniteLinearChain}). By contrast, for a finite chain of length $N$, at nucleus $l$ the coupling parameter $K_l$ is given by 
\begin{align}
 K_l  &=  \sum_{m=1}^N \mathcal{C}^*_{lm} \: (e^{i\,m\,\Delta\phi} + e^{-i\,m\,\Delta \phi}) \nonumber \\
 &=-  3 i \Gamma^\mathrm{rad}  \sum_{m=1}^\infty \left[ \left(\frac{1}{\eta_0 m}+\frac{i}{\eta_0^2 m^2}-\frac{1}{\eta_0^3 m^3}\right) \right. \nonumber \\
 &\left. \quad - \cos^2(\theta_d)\: \left(\frac{1}{\eta_0 m}+\frac{3i}{\eta_0^2 m^2}-\frac{3}{\eta_0^3 m^3} \right) \right]\: e^{i\,m\,\eta_0}\nonumber \\
 &\quad \times (e^{i\,m\,\Delta\phi} + e^{-i\,m\,\Delta \phi}) \,.
\end{align}
In Fig.~\ref{fig:convergenceOfK}(a), finite-size effects in the coupling parameter $K$ are explored. It compares the converged $K$ for an infinite chain  with the corresponding parameter for the central nucleus $l=N/2$ in a finite chain of variable length $N$. Overall, we observe a convergence to the  asymptotic value with increasing $N$. However, the convergence is not monotonic, but governed by oscillations around the asymptotic value with frequency decreasing for smaller incident light angles $\theta_\mathrm{in}$. The effect of this dependence could be explored in small nuclear ensembles.

Fig.~\ref{fig:convergenceOfK}(b) shows the coupling parameter for a finite chain $K_l$ in Eq.~(\ref{eq:K_finite})  evaluated as function of the position within the chain $l$. A pronounced oscillatory dependence on $l$ is clearly visible, depending on the asymmetry in the number of nuclei to the  left and right of a certain nucleus $l$. The oscillations decrease with rising number of nuclei $N$. However, for smaller chains, they are  of relevance in evaluating the far-field scattered electric field, and may modify the angular pattern already in the  low-excitation limit~\cite{PhysRevA.90.063834}. For larger samples, the variations are negligible, which is why we continue with the analysis of finite size effects at the central nucleus only.

To investigate finite size effects during the time evolution, in Fig.~\ref{fig:phase_finiteSizeEffects} the results for the central nucleus in a finite chain (dashed lines) are compared to those from the translational invariant model (dotted lines) for different initial excitations $\mathcal{A}$. For Fig.~\ref{fig:phase_finiteSizeEffects}(a) the number of nuclei is $N=362$, which corresponds to a maximal difference of the coupling parameter $K^I$ from the asymptotic value as can be seen from Fig.~\ref{fig:convergenceOfK}. Although qualitatively the overall trends of the non-linear time-evolution agree for the finite and infinite chain, there are noticeable quantitative differences. In Fig.~\ref{fig:phase_finiteSizeEffects}(b) a smaller chain with $N=257$ nuclei, but with minimal deviation of the coupling parameter $K^I$ is used. Here, the quantitative agreement is significantly better than in the other case although the chain is smaller. This is already a strong indication, that the dominant finite size effect for observables on the single atom level is the convergence of the coupling parameter $K$.

\begin{figure}
    \centering
    \includegraphics[width=\linewidth]{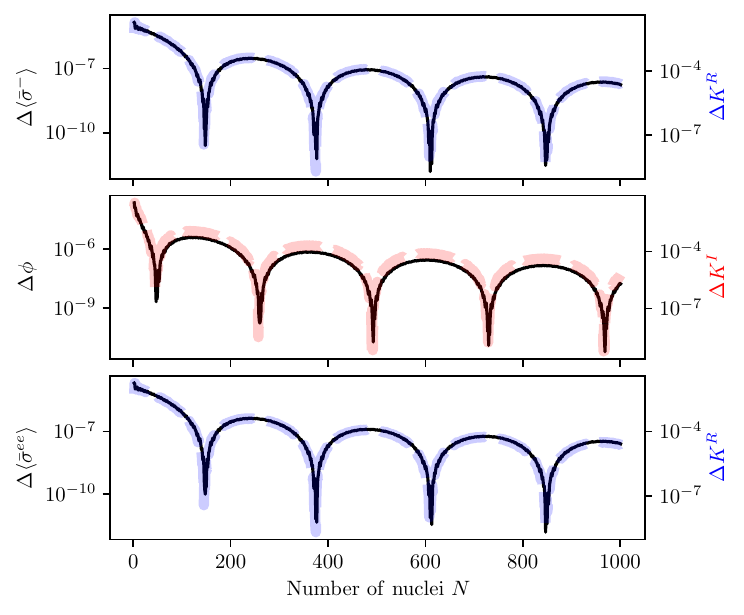}
    \caption{Impact of finite size effects on the nuclear dynamics. The deviation of observable $O$ is quantified  by the integrated quadratic differences over  time $\Delta O = \frac{1}{M}\sum_{i=1}^{M}|O_\infty(t_i)-O_N(t_i)|^2$, where $O_\infty$ is the result for a  translationally invariant  infinite chain and $O_N$ is the result from a finite chain of length $N$ evaluated at the central nucleus. The three panels show the deviation as function of chain length $N$ (black markers). The comparison is made for the absolute value of the nuclear coherence (top panel), the phase of the coherence (central panel) and the population (bottom). For comparison, on the second y-axis, the squared absolute difference $\Delta K=|K_\infty - K_N|^2$ of the dominant part (real or imaginary part) of the coupling parameter $K$ is shown (red and blue dashed lines). The integration time is up to $t=2/\Gamma$, the dipole is perpendicular to the chain $\theta_d=\pi/2$, the incident light angle is $\theta_\mathrm{in}=50\,$mrad and the initial excitation is $\mathcal{A}=\pi/2$. }
    \label{fig:convergence_with_N}
\end{figure}

To quantify the convergence with chain length, for each observable $O$ we calculate the mean squared absolute difference between the result from the translational invariant model $O_\infty$ and the finite chain of length $N$ result $O_N$ integrated over time $\Delta O = \frac{1}{M}\sum_{i=1}^{M}|O_\infty(t_i)-O_N(t_i)|^2$. In Fig.~\ref{fig:convergence_with_N} the deviation $\Delta O$ is displayed for the absolute value and phase of the coherence and the population evaluated at the central nucleus in the chain $l=N/2$ as function of the number of nuclei in the chain $N$. The general trend is that the deviations decrease with rising length of the chain as expected. In addition, the deviations for the absolute values of the coherence and the population are smaller because in the respective equations of motions, the relevant part of the coupling parameter $K^R$ is dominated by the single-particle decay $\Gamma$. Besides the general convergence, there is an oscillatory structure. This structure comes from the oscillatory convergence behavior as shown in Fig.~\ref{fig:convergenceOfK}. To see this more clearly, on a second y-axis for the real and imaginary part, respectively, the squared absolute difference of the coupling parameter calculated for a finite chain $K$ is compared to the asymptotic value given by Eq.~\ref{eq:K_infiniteLinearChain} is displayed. The oscillatory structure of $\Delta O$ clearly follows the differences of $K$. Therefore, we conclude that for the central nucleus the most-dominant finite size effect is the difference in the imaginary part of the coupling parameter $K^I$, which in a converging but oscillating manner depends on the number of nuclei in the chain $N$.

\begin{figure}
    \centering
    \includegraphics[width=\linewidth]{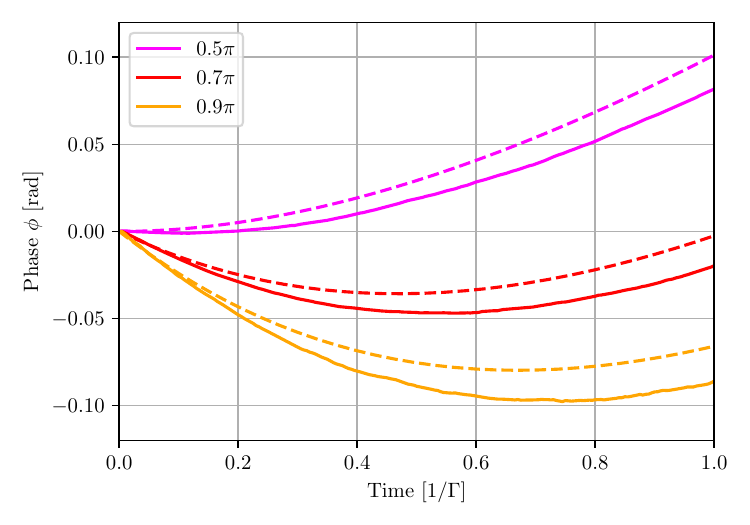}
    \caption{Phase evolution of the central nucleus in a chain of length $N=100$. The figure compares results of the  cumulant expansion (dashed) with corresponding results obtained from the  truncated Wigner approximation (solid). The different colors indicate different initial excitations $\mathcal{A}$. The calculations are performed without internal conversion ($\Gamma^\mathrm{IC}=0$) for dipole moment perpendicular to the chain ($\theta_d=\pi/2$) and a small incident light angle of $\theta_\mathrm{in}=5$\,mrad.}
    \label{fig:cumulant_vs_TWA}
\end{figure}

\textit{Beyond cumulant expansion.}
In order to verify that the results discussed in the main text obtained from the cumulant expansion are within the validity range of the method, we compare the results with the phase evolution obtained in the framework of the continuous-discrete truncated Wigner approximation (CDTWA) developed by Mink et al. \cite{Mink2023}. 
This approach is known to work best for stronger couplings and higher excitations, whereas in the opposite case  beyond $t=1/\Gamma$ unphysical effects appear~\cite{Mink2023}. Therefore, we neglect the internal conversion for this comparison  ($\Gamma^\mathrm{IC}=0$) which results in higher cooperativity and thus better validity of CDTWA, but  is a less favorable parameter range for the cumulant expansion. With internal conversion, the cumulant expansion results are expected to be be more reliable.  Nonetheless,  the CDTWA provides a good crosscheck for the cumulant expansion results because the approximation methods are complementary to each other. 
In Fig.~\ref{fig:cumulant_vs_TWA} for the central nucleus $l=N/2$ in a chain of length $N=100$, the phase evolution calculated with the cumulant expansion and the CDTWA is compared for different initial excitations $\mathcal{A}$. For all three shown different degrees of initial excitation, the results obtained with the cumulant expansion (dashed lines) and CDTWA (solid lines) agree qualitatively well, which confirms the  non-linear phase evolution.

\begin{figure}
    \centering
    \includegraphics[width=0.6\linewidth]{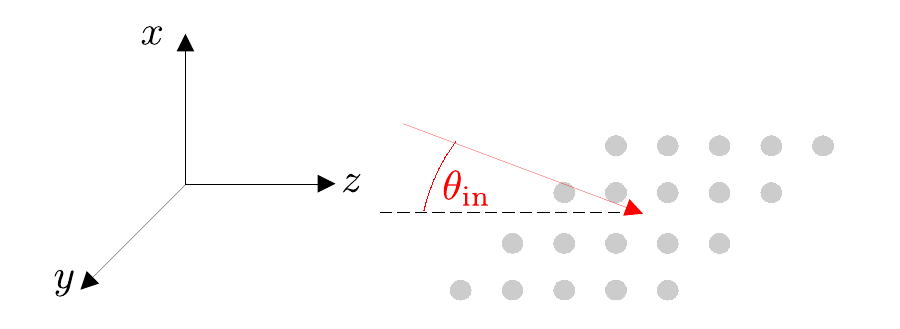}
    \caption{Geometry of the 2D setup. The 2D array of nuclei (gray) is placed in the $yz$-plane. The incident light beam (red) lies in the $xz$-plane with an incident light angle $\theta_\mathrm{in}$ relative to the 2D array.}
    \label{fig:2D_geometry}
\end{figure}

\begin{figure*}
    \centering
    \includegraphics[width=0.9\linewidth]{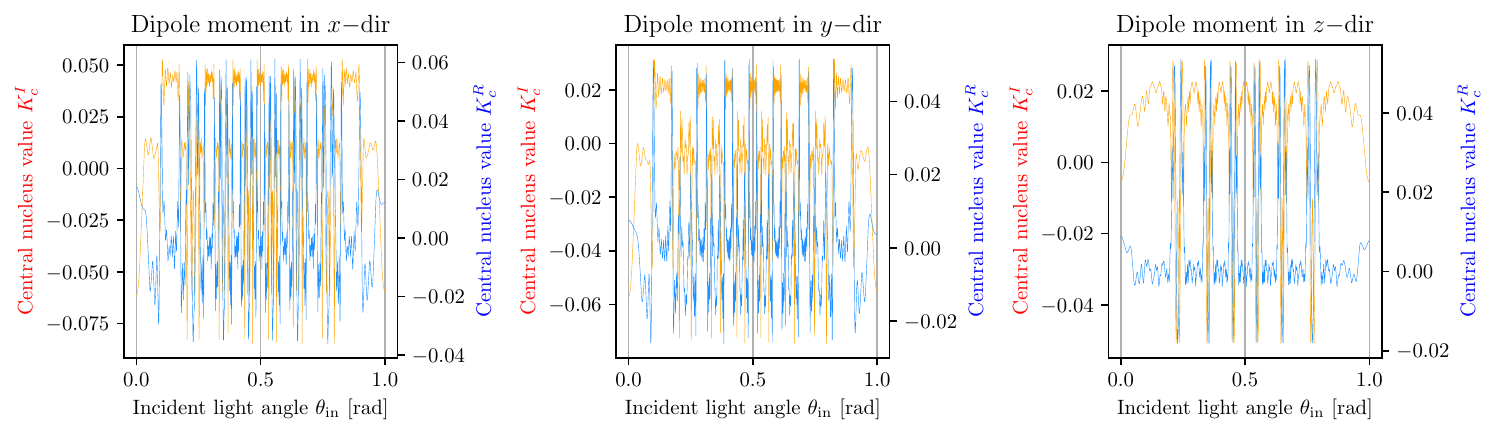}
    \includegraphics[width=0.9\linewidth]{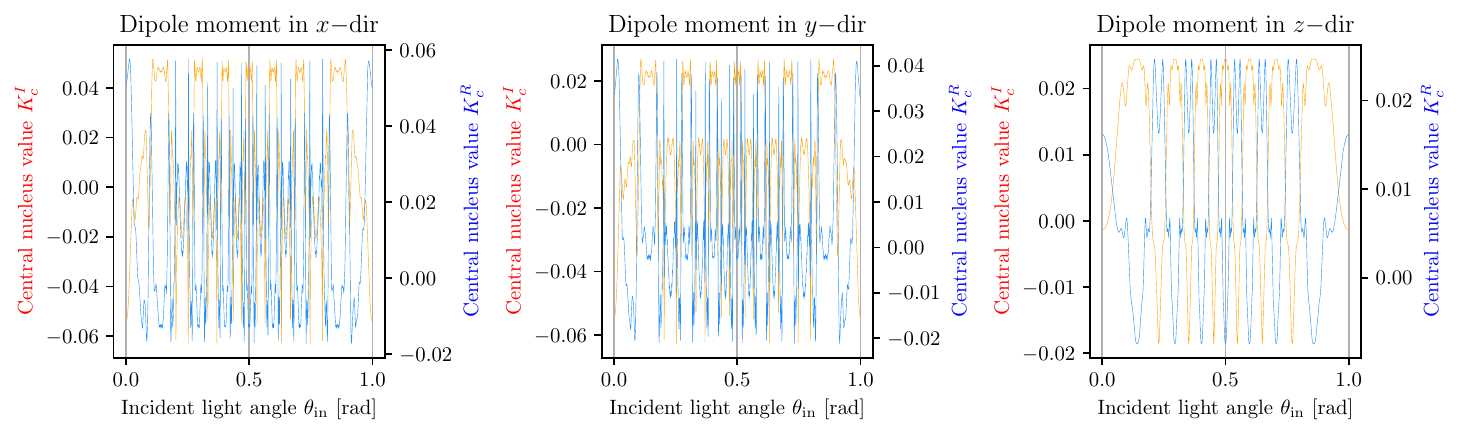}
    \caption{Coupling parameter $K$ as function of the incident light angle $\theta_\textrm{in}$ of a nucleus in the center of the plane. The different rows show a square array with $50\times50$ nuclei (top) and a rectangular array with $10\times100$ nuclei elongated in $z$-direction (bottom). The three columns correspond to the dipole moment being aligned in the three directions.}
    \label{fig:K_in2D}
\end{figure*}

\begin{figure}
    \centering
    \includegraphics[width=0.9\linewidth]{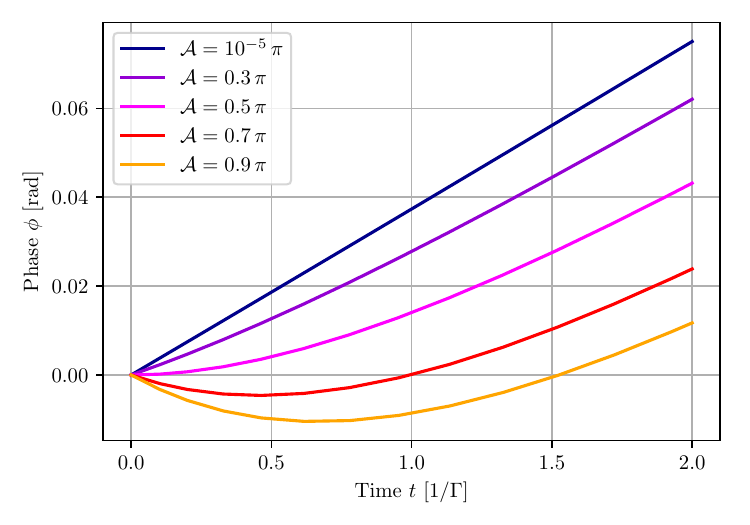}
    \caption{Phase evolution of central nucleus as function of time for different excitations indicated by the different colors. The grid is $50\times50$, the dipole moment is chosen along the $y$-direction and the incident light angle $\theta_\textrm{in}=5\,\text{mrad}$.}
    \label{fig:2D_dynamics}
\end{figure}

\textit{Generalization to  2D geometries.}
In the main part we discussed a 1D chain along the z-axis as the simplest model. Here, we generalize the analysis to 2D arrays. A schematic of the geometry of the 2D extension is shown in Fig.~\ref{fig:2D_geometry}. The 2D array of nuclei lies in the $yz$-plane. The incident light propagates in the $xz$-plane and impinges onto the plane in which the nuclei are located under an angle $\theta_\mathrm{in}$. 

First, we study the dependence of the coupling parameter $K$ on the incident light angle $\theta_\mathrm{in}$. 
In Fig.~\ref{fig:K_in2D} we show $K$ as function of the incident light angle $\theta_\textrm{in}$ for different settings. The top row shows the result for a $50\times50$ square lattice, and the bottom row for a $10\times100$ lattice elongated in $z$-direction. Both shapes can  in principle be produced, but the square lattice is closer to the idea of a translationally invariant system in 2D. The three columns show the results of all dipole-moments aligned along the $x$- (left), $y$- (center) and $z$-direction (right). 

As a first result, we find that the order of magnitude that is reachable for $K$ is similar in all arrangements, except for small angles where in $z$-direction it is less than in the 1D case. Interestingly, the square array and the rectangular array with dipoles aligned in $x$- or $y$-directions give quite similar results over a broad range of incidence angles. 

For small angles, having the dipole moment aligned in $x$-direction (perpendicular to plane) results in larger values of $K^R_\textrm{center}$ than in $y$-direction (in plane) while $K^I_\textrm{center}$ is approximately the same. The difference between those two directions is important because experimentally the alignment of  the dipole moments out of plane is more challenging. Another important observation is that $K^R_\textrm{center}$ also has a rich structure as function of the incident angle $\theta_\textrm{in}$ and is not constant at small angles. Nonetheless, as in the 1D case, tuning the incident angle $\theta_\textrm{in}$ allows one to tune to $K^R_\textrm{center}=0$, which is crucial for the proposed interferometric phase measurement.

For an incident light angle of $\theta_\textrm{in}=5\,\text{mrad}$ as for the investigations in the main text, a dipole moment in $y$-direction and the $50\times50$ array, we find 
\begin{align}
    K^I_\textrm{center} \approx &  -0.056\,,\\
    K^R_\textrm{center} \approx & +0.0074  \, .
\end{align}
Thus, $K^I_\textrm{center}$ (responsible for the phase evolution) is on a similar order as in the 1D case, while $K^R_\textrm{center}$ (responsible for the decay) is even smaller than in the 1D case.
For the $10\times100$ array, we find 
\begin{align}
    K^I_\textrm{center} \approx &  -0.053\,,\\
    K^R_\textrm{center} \approx & +0.035  \, .
\end{align}
In this case, $K^I_\textrm{center}$ is on a similar order as in the 1D case, while $K^R_\textrm{center}$ is approximately three times higher than in the 1D case, however small compared to $\Gamma$.

After studying the coupling parameter $K$, we can now study the phase evolution of a central nucleus for a fixed geometry. In Fig.~\ref{fig:2D_dynamics} this phase evolution is exemplarily shown for a $50\times50$ array with the dipole moments along the $y$-direction and an incident light angle of $\theta_\textrm{in}=5\,\text{mrad}$ as in the main text. The respective colors represent different degrees of excitation $\mathcal{A}$. While quantitatively the scale of the phase evolution is slightly smaller than in the 1D case, qualitatively the behavior is  similar. At low degree of excitation, the phase evolution has a linear slope, which can be associated with a detuning. For higher excitations, the phase evolution becomes non-linear.

Overall, we therefore conclude that the key results reported in the main text generalize to 2D configurations in suitably chosen scattering geometries. Most importantly, the non-linear phase evolution is present, on a  quantitatively similar scale than in the 1D case studied in the main text.

\noindent
\textit{Details on the coupling parameter $K$.}
To describe the coupling parameter $K$ as a function of the geometry of the nuclear ensemble, we assume that the nuclear dipole moments are aligned at an angle $\theta_d$ relative to the chain axis, and that the incident plane wave x-ray pulse propagates at an angle $\theta_\mathrm{in}$ with respect to the chain. We model the couplings using the well-known expression for free-space dipole-dipole couplings~\cite{Ficek_Swain,Kiffner_Vacuum_Processes}
\begin{align} \label{eq:transInv_couplingsGeneral}
 J_{nm} + i \frac{\Gamma_{nm}}{2} &= \frac{1}{\hbar} \: \vec{d}^*\cdot \chi(\vec{r}_n, \vec{r}_m ) \cdot \vec{d}\,,
\end{align}
with
\begin{align}
  \chi_{pq}(\vec{r}_n, &\vec{r}_m )= \frac{k_0^3}{4\pi\varepsilon_0} \left[\delta_{pq}\left(\frac{1}{\eta_{nm}}+\frac{i}{\eta_{nm}^2}-\frac{1}{\eta_{nm}^3}\right) \right. \nonumber \\
  &\left. - \frac{[\vec{R}_{nm}]_p\, [\vec{R}_{nm}]_q}{|\vec{R}_{nm}|^2}\left(\frac{1}{\eta_{nm}}+\frac{3i}{\eta_{nm}^2}-\frac{3}{\eta_{nm}^3} \right) \right]\: e^{i\eta_{nm}}\,.
\end{align}
Here it is, $\vec{R}_{nm} = \vec{r}_n -\vec{r}_m$ and $\eta_{nm} = k_0 |\vec{R}_{nm}|$.

We continue the discussion for a linear chain of nuclei with a dipole moment at angle $\theta_d$ and the incident light at angle $\theta_\mathrm{in}$ with respect to the chain. For a linear chain of atoms along the z-direction, we can set $\vec{r}_n = (0,0, n\cdot a_0)^\intercal$, where $a_0$ is the lattice constant. Then, $\eta_{nm} = k_0 a_0 |n-m| = \eta_0 |n-m|$, and Eq.~\ref{eq:transInv_couplingsGeneral} simplifies to
\begin{align}
 &J_{nm} + i \frac{\Gamma_{nm}}{2} =  \frac{k_0^3 |d|^2}{4\pi\hbar \varepsilon_0} \left[ \left(\frac{1}{\eta_{nm}}+\frac{i}{\eta_{nm}^2}-\frac{1}{\eta_{nm}^3}\right) \right. \nonumber \\
 &- \left. \cos^2(\theta_d)\: \left(\frac{1}{\eta_{nm}}+\frac{3i}{\eta_{nm}^2}-\frac{3}{\eta_{nm}^3} \right) \right]\: e^{i\eta_{nm}}\,,
 \end{align}
where $\theta_d$ is the orientation of the dipole moment relative to the chain. We thus find
\begin{align} \label{eq:C_finite_supp}
 \mathcal{C}^*_{nm}& = - \frac{3 i \Gamma^\mathrm{rad}}{2} \left[ \left(\frac{1}{\eta_{nm}}+\frac{i}{\eta_{nm}^2}-\frac{1}{\eta_{nm}^3}\right) \right. \nonumber \\
 &\left.- \cos^2(\theta_d)\: \left(\frac{1}{\eta_{nm}}+\frac{3i}{\eta_{nm}^2}-\frac{3}{\eta_{nm}^3} \right) \right]\: e^{i\eta_{nm}}\,,
\end{align}
with the radiative decay rate $\Gamma^\mathrm{rad}=\Gamma^\mathrm{rad}_{nn}$, which is significantly smaller than the total decay rate $\Gamma$.
Therefore, for an infinite chain of nuclei with free-space coupling, the coupling parameter $K$ is given by
\begin{align}
 K 
 = &-  3 i \Gamma^\mathrm{rad}  \sum_{m=1}^\infty \left[ \left(\frac{1}{\eta_0 m}+\frac{i}{\eta_0^2 m^2}-\frac{1}{\eta_0^3 m^3}\right) \right. \nonumber \\
  & \left. - \cos^2(\theta_d)\: \left(\frac{1}{\eta_0 m}+\frac{3i}{\eta_0^2 m^2}-\frac{3}{\eta_0^3 m^3} \right) \right]\: e^{i\,m\,\eta_0} \nonumber \\ 
  &\quad  \times (e^{i\,m\,\Delta\phi} + e^{-i\,m\,\Delta \phi}) \, .
\end{align}
This expression can be simplified using the polylogarithm $\sum_{k=1}^\infty \frac{z^k}{k^n}=\mathrm{Li}_n(z)$ if $z\neq1$, and we find 
\begin{align}
 K =&  \textcolor{black}{-}\frac{3i\Gamma^\mathrm{rad}}{2 \eta_0} \textcolor{black}{\sin^2(\theta_d)} \:\mathcal{X}_1 
\textcolor{black}{-}\frac{3\Gamma^\mathrm{rad}}{4\eta_0^2}[1+3\cos(2\theta_d)] \:\mathcal{X}_2  \nonumber \\
&\textcolor{black}{-}\frac{3i\Gamma^\mathrm{rad}}{4\eta_0^3}[1+3\cos(2\theta_d)]\: \mathcal{X}_3 \,,\nonumber \\[2ex]
 \mathcal{X}_n =& \mathrm{Li}_n\left( e^{i(\eta_0 +\Delta \phi)}\right) + \mathrm{Li}_n\left( e^{i(\eta_0 -\Delta \phi)}\right)\,.
\end{align}
For the numerical analysis, we use wavevector $k_0=2\pi/86\,\text{pm}$ corresponding to the resonance frequency of the M\"ossbauer transition in ${}^{57}$Fe and a lattice constant $a_0=286\,$pm corresponding to the nearest-neighbor spacing in $\alpha$-Fe. With these parameters,  the scaled distance parameter $\eta_0=k_0 a_0\approx 21$ reflects the weak coupling due to the comparably large separation of the nuclei, justifying the cluster expansion approach. Further, $\Delta \phi = \eta_0 \cos \theta_\mathrm{in}$ characterizes the incident phase difference between two neighboring nuclei.

The dipole-dipole coupling is mediated by virtual photon exchange. As compared to free space, the photon propagation is modified by  the bulk material. In particular, absorption leads to a finite range of the coupling. To include these effects, we start from Eq.~\ref{eq:C_finite_supp}, and replace the vacuum wavevector $k_0$ with the one in the medium
\begin{align}
    k_0 -\rightarrow k' = nk_0 = (1-\delta + i\beta)k_0 \, ,
\end{align}
where $n$ is the refractive index. For  iron at 14.4~keV photon energy,  the corrections $\delta=7.4\times 10^{-6}$, $\beta=3.4\times 10^{-7}$ from $n=1$ are small~\cite{Henke1993}. 

\begin{figure}
    \centering
    \includegraphics[width=0.8\linewidth]{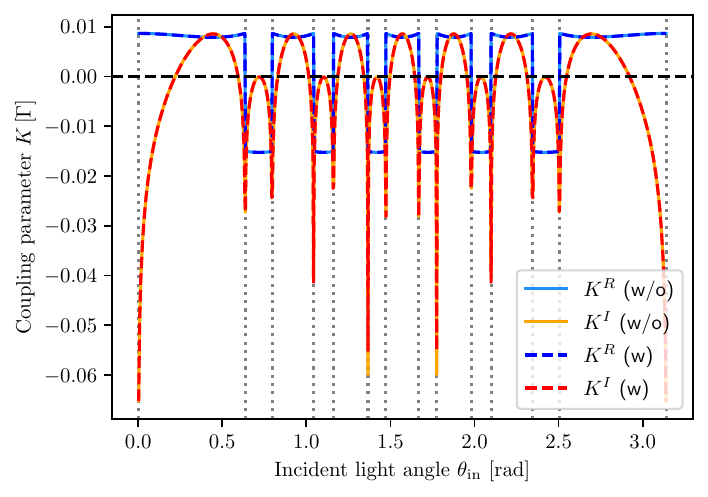}
    \caption{Comparison of $K$ without (solid) and with (dashed lines) considering absorption in the medium via the modified wavevector $k$ in all terms. The equation of the asymptotic value for $K$ is used. The gray doted lines indicate the positions, where the polylogarithm diverges. The dipole orientation is chosen to be perpendicular to the chain ($\theta_d=\pi/2$).}
    \label{fig:K_withAbsorption}
\end{figure}

Considering the structure of the coupling constant, the dominant effect of the refractive index arises from the absorption in the phase entering the exponential function. Taking into account this correction to the phase modifies the couplings  to
\begin{align}
 \mathcal{C}'^{*}_{nm}& = - \frac{3 i \Gamma^\mathrm{rad}}{2} \left[ \left(\frac{1}{\eta_{nm}}+\frac{i}{\eta_{nm}^2}-\frac{1}{\eta_{nm}^3}\right) \right. \nonumber \\
 &\left. \quad - \cos^2(\theta_d)\: \left(\frac{1}{\eta_{nm}}+\frac{3i}{\eta_{nm}^2}-\frac{3}{\eta_{nm}^3} \right) \right]\: e^{i\eta_{nm}} e^{-\beta \eta_{nm}}\,.
 \end{align}
As in the vacuum case, the coupling parameter $K$ in an infinite chain is given by
\begin{align}
 K' &=  \sum_{m=1}^\infty \mathcal{C}'^*_{0m} \: (e^{i\,m\,\Delta\phi} + e^{-i\,m\,\Delta \phi}) \nonumber \\
 &=-  3 i \Gamma^\mathrm{rad}  \sum_{m=1}^\infty \left[ \left(\frac{1}{\eta_0 m}+\frac{i}{\eta_0^2 m^2}-\frac{1}{\eta_0^3 m^3}\right) \right. \nonumber \\
 &\left. \quad - \cos^2(\theta)\: \left(\frac{1}{\eta_0 m}+\frac{3i}{\eta_0^2 m^2}-\frac{3}{\eta_0^3 m^3} \right) \right]\: e^{i\,m\,\eta_0} e^{-\beta m \eta_{0}}\nonumber \\
 &\quad \times (e^{i\,m\,\Delta\phi} + e^{-i\,m\,\Delta \phi}) \nonumber   \\[2ex]
  &= -\frac{3i\Gamma^\mathrm{rad}}{2 \eta_0} \sin^2(\theta) \:\mathcal{X}_1' 
-\frac{3\Gamma^\mathrm{rad}}{4\eta_0^2}[1+3\cos(2\theta)] \:\mathcal{X}_2' \nonumber \\
&\quad -\frac{3i\Gamma^\mathrm{rad}}{4\eta_0^3}[1+3\cos(2\theta)]\: \mathcal{X}_3' \,, 
\end{align}
where
\begin{align}
 \mathcal{X}_n' &= \mathrm{Li}_n\left( e^{i(\eta_0 +\Delta \phi)-\beta\eta_0}\right) + \mathrm{Li}_n\left( e^{i(\eta_0 -\Delta \phi)-\beta\eta_0}\right)\,. \label{eq:K_refIndex}
\end{align}
Due to the $-\beta \eta_0$ contribution, the arguments of the polylogarithm $\mathrm{Li}_n(z)$ are always $|z|<1$ and therefore the polylagorithm is always defined. This way,  singularities appearing at selected incident angles in the non-absorbing case are removed by the absorption. In Fig.~\ref{fig:K_withAbsorption} we show the comparison with and without consideration of the refractive index. Indeed, the deviations can only be seen  where the singularities due to the polylogarithm (gray dotted lines) appear in the non-absorbing case. Note that in this comparison, the refractive index enters the wavevector $k$ and is thus taken into account everywhere in the coupling including the $\eta_{nm }$ in the denominators, and not only approximately as in Eq.~\ref{eq:K_refIndex}.

\noindent
\textit{First vs. second order cumulant expansion.}
It is known that in the first order cumulant expansion certain effects, e.g. phase synchronization, might not be captured~\cite{photonics12100996,Ostermann2019}. For this reason, in the Appendix we compare our results to that of the continuous-discrete truncated Wigner approximation~\cite{Mink2023}, which is a complementary method for the simulation of may-body ensembles. In addition, in the following we  compare the results in the main text obtained with the first order cumulant expansion to corresponding results obtained with the second order cumulant expansion. To this end, we implement our system using the julia package \texttt{QuantumCumulants.jl}~\cite{plankensteiner2022quantumcumulants}, which in principle allows one to calculate the dynamics to arbitrary order of cumulant expansion.

\begin{figure}
    \centering
    \includegraphics[width=0.9\linewidth]{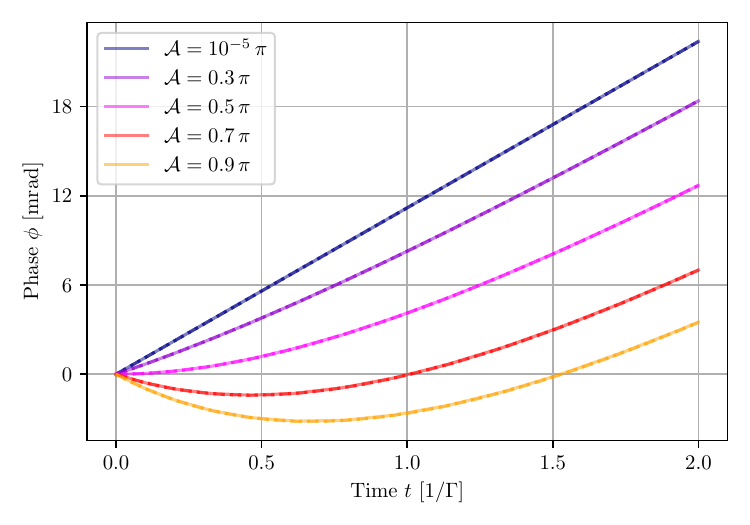}
    \caption{Comparison of the phase evolution for different methods (line styles) and initial excitation (colors). The used methods are our own implementation of the first order (solid) and the \texttt{QuantumCumulants} julia-package in first order (dotted) and second order (dashed). The chain of $N=10$ nuclei is along the $z$-direction, the dipole moment parallel to the chain and the incident light angle $\theta_\mathrm{in}=5\text{mrad}$ as in the main text. }
    \label{fig:1st_vs_2nd_order}
\end{figure}

In Fig.~\ref{fig:1st_vs_2nd_order}, we compare the first and second order results of the phase evolution for a chain of $N=10$ nuclei. The dipole moments are aligned perpendicular to the chain and the incident light angle is $\theta_\mathrm{in}=5\,\text{mrad}$ as in the main text. The phase evolution for different excitations (colors) and first (solid) and second order (dashed) is displayed. We can clearly see that the results from the first order and the second order agree well with each other. 
The comparison between first and second order cumulant expansion provides an additional cross-check whether the important physics is captured. Due to the rather weak couplings and the synchronization of the nuclei due to the exciting x-ray pulse, the relevant physics for the discussed dipole phase evolution is already captured in the first order cumulant expansion.

\textit{Defects and disorder.}
In this section, we study how imperfections in form of defects and disorder influence the results for an ideal chain.

\begin{figure*}
    \centering
    \includegraphics[width=0.9\linewidth]{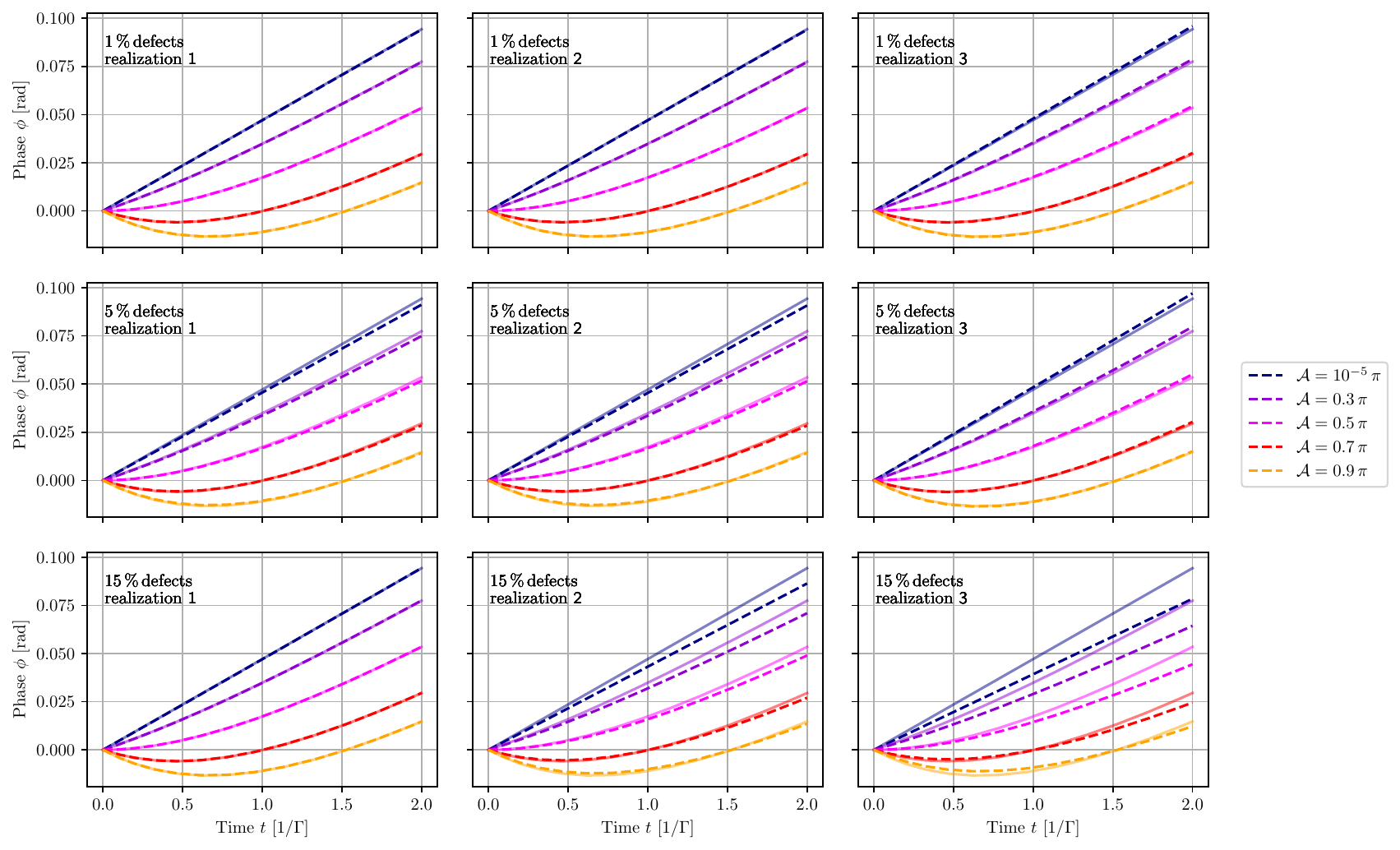}
    \caption{Time evolution of the dipole phase of the central nucleus in an ideal chain (solid lines) and a chain with defects (dashed lines) for different initial excitations $\mathcal{A}$ (different colors). Each row shows three realizations of the random defects, i.e., the empty lattice positions determined  according to the specified defect rate. The chain of $N=1000$ nuclei is along the $z$-direction, the dipole moment parallel to the chain and the incident light angle $\theta_\mathrm{in}=5\text{mrad}$ as in the main text.}
    \label{fig:defects_examples}
\end{figure*}

Typical, ${}^{57}$Fe samples are enriched to approximately 95\% of ${}^{57}$Fe while the remaining 5\% are ${}^{56}$Fe, which does not have a Mössbauer transition but the same electronic properties. Therefore, we can treat this imperfection as defects, i.e. empty positions, in the lattice. 
 
Those positions are randomly chosen according to a specified defect rate. In Fig.~\ref{fig:defects_examples}, a chain of length \textcolor{black}{N=1000} is considered with a dipole moment perpendicular to the chain and an incident light angle of $\theta_\mathrm{in}=5~\text{mrad}$ as in the main text. For different defect rates (different rows) we  then compare the dipole phase evolution in an ideal chain (solid lines) to that of a chain with defects (dashed lines). Note that in an experiment, each sample would correspond to a random, but fixed realization of the defects.
In general, the larger the defect rate, the stronger the deviation from the result for an ideal chain. As expected this varies for different realizations of a certain defect rate. Typically the deviations lead to a slightly smaller time evolution of the phase because empty positions in the lattice lead to reduced dipole-dipole couplings. However, in all cases, the general behavior is the same as for the ideal chain: There is a non-linear phase evolution that clearly depends on the initial degree of excitation $\mathcal{A}$. In addition, in the experimentally relevant case of a defect rate of 5\%, the results are still close to that of an ideal chain.

To summarize, even in an imperfect chain with empty positions, the non-linear dipole phase evolution and especially the dependence on the initial degree of excitation $\mathcal{A}$ are clearly visible.

\begin{figure*}
    \centering
    \includegraphics[width=0.9\linewidth]{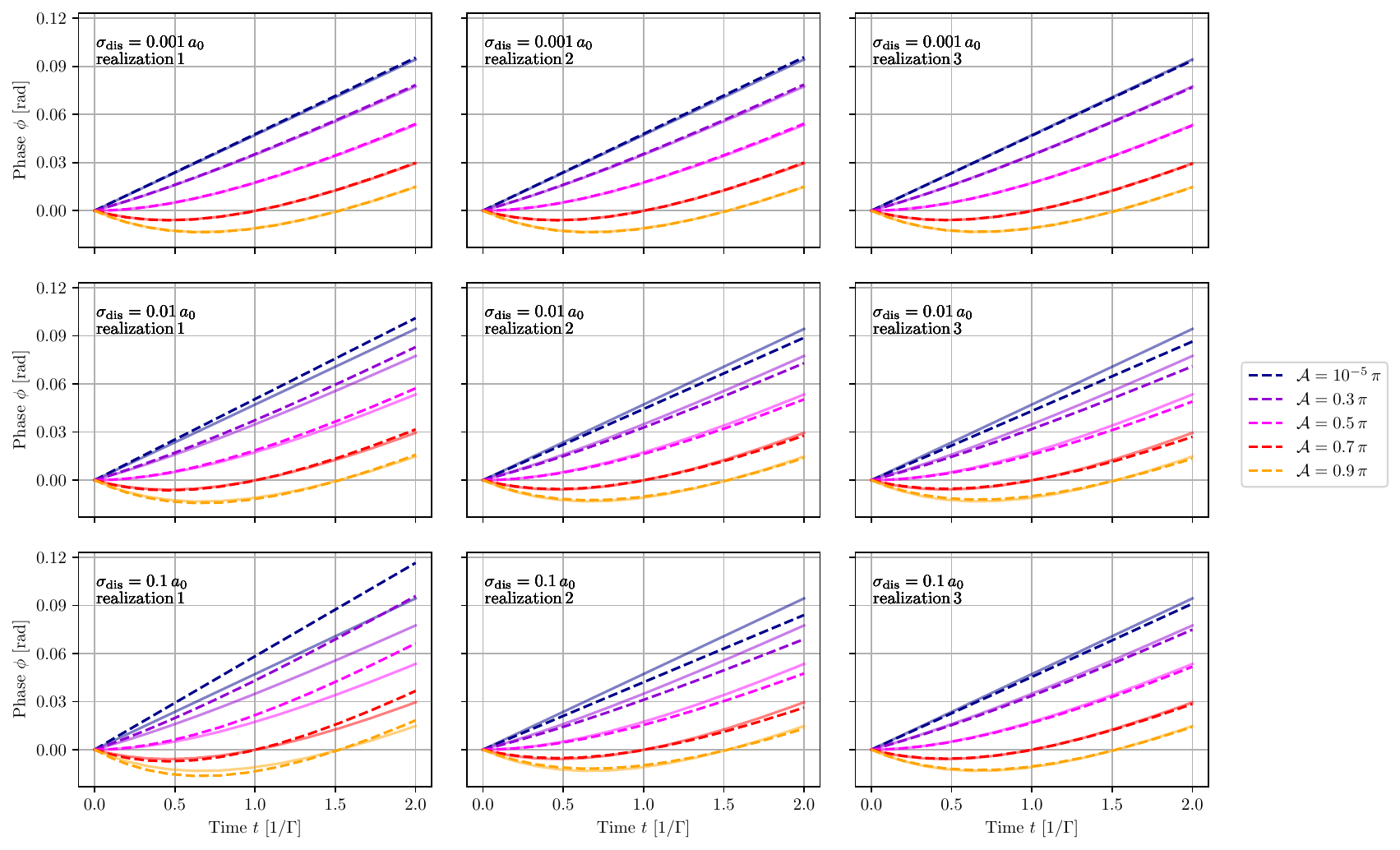}
    \caption{Time evolution of the dipole phase of the central nucleus in an ideal chain (solid lines) and a chain with imperfections in the lattice (dashed lines) for different initial excitations $\mathcal{A}$ (different colors). Each rows shows three realizations of random displacements drawn for each nuclei from a normal distribution centered around zero with different standard deviations $\sigma_\textrm{dis}$, where $a_0$ is the lattice constant. The chain of $N=1000$ nuclei is along the $z$-direction, the dipole moment parallel to the chain and the incident light angle $\theta_\mathrm{in}=5~\text{mrad}$ as in the main text.}
    \label{fig:disorder_examples}
\end{figure*}

Besides defects, another kind of chain imperfection  is disorder. Although the nuclei are embedded in a solid state lattice and the positions are expected to be on the lattice grid, next we quantify  the influence of disorder.

The disorder is implemented by shifting each nucleus separately from its equidistant lattice positions by a random number drawn from a normal distribution centered around zero with a standard deviation of $\sigma_\textrm{dis}$. In Fig.~\ref{fig:disorder_examples} we compare the phase evolution of the central nucleus in an ideal chain (solid lines) and a chain with disorder (dashed lines). Again, we consider a chain of length \textcolor{black}{N=1000}   with a dipole moment perpendicular to the chain and an incident light angle of $\theta_\mathrm{in}=5~\text{mrad}$ as in the main text. The different rows show three realizations of randomly drawn disorder for different width $\sigma_\textrm{dis}$ of the normal distribution. As in the case of defects, the deviations from the ideal chain grow with increasing disorder, and the results in the different realizations vary. Nonetheless, in all cases the non-linear phase evolution and its dependency on the initial degree of excitation $\mathcal{A}$ are clearly visible.

Thus, we can again conclude that also disorder does not qualitatively change the results obtained for an ideal chain.

\newpage 

\bibliography{literature}

@Article{Mink2023,
	title={{Collective radiative interactions in the discrete truncated Wigner approximation}},
	author={Christopher D. Mink and Michael Fleischhauer},
	journal={SciPost Phys.},
	volume={15},
	pages={233},
	year={2023},
	publisher={SciPost},
	doi={10.21468/SciPostPhys.15.6.233},
	url={https://scipost.org/10.21468/SciPostPhys.15.6.233},
}

@article{lukas,
  title = {Characterization and detection method for x-ray excitation of {M}\"ossbauer nuclei beyond the low-excitation regime},
  author = {Wolff, Lukas and Evers, J\"org},
  journal = {Phys. Rev. A},
  volume = {108},
  issue = {4},
  pages = {043714},
  numpages = {20},
  year = {2023},
  month = {Oct},
  publisher = {American Physical Society},
  doi = {10.1103/PhysRevA.108.043714},
  url = {https://link.aps.org/doi/10.1103/PhysRevA.108.043714}
}

@article{kubo1962generalized,
  title={Generalized cumulant expansion method},
  author={Kubo, Ryogo},
  journal={Journal of the Physical Society of Japan},
  volume={17},
  number={7},
  pages={1100--1120},
  year={1962},
  publisher={The Physical Society of Japan}
}

@article{Longo2016,
  title = {Tailoring superradiance to design artificial quantum systems},
  volume = {6},
  ISSN = {2045-2322},
  DOI = {10.1038/srep23628},
  number = {1},
  journal = {Scientific Reports},
  publisher = {Springer Science and Business Media LLC},
  author = {Longo,  Paolo and Keitel,  Christoph H. and Evers,  J\"{o}rg},
  year = {2016},
  month = mar 
}

@article{oliver,
  title = {Inverse design approach to x-ray quantum optics with {M}\"ossbauer nuclei in thin-film cavities},
  author = {Diekmann, Oliver and Lentrodt, Dominik and Evers, J\"org},
  journal = {Phys. Rev. A},
  volume = {105},
  issue = {1},
  pages = {013715},
  numpages = {18},
  year = {2022},
  month = {Jan},
  publisher = {American Physical Society},
  doi = {10.1103/PhysRevA.105.013715},
  url = {https://link.aps.org/doi/10.1103/PhysRevA.105.013715}
}

@article{PhysRevA.106.053701,
  title = {Inverse design in nuclear quantum optics: From artificial x-ray multilevel schemes to spectral observables},
  author = {Diekmann, Oliver and Lentrodt, Dominik and Evers, J\"org},
  journal = {Phys. Rev. A},
  volume = {106},
  issue = {5},
  pages = {053701},
  numpages = {25},
  year = {2022},
  month = {Nov},
  publisher = {American Physical Society},
  doi = {10.1103/PhysRevA.106.053701},
}

@article{PhysRevA.90.063834,
  title = {Probing few-excitation eigenstates of interacting atoms on a lattice by observing their collective light emission in the far field},
  author = {Longo, Paolo and Evers, J\"org},
  journal = {Phys. Rev. A},
  volume = {90},
  issue = {6},
  pages = {063834},
  numpages = {27},
  year = {2014},
  month = {Dec},
  publisher = {American Physical Society},
  doi = {10.1103/PhysRevA.90.063834},
  url = {https://link.aps.org/doi/10.1103/PhysRevA.90.063834}
}

@article{xiangjin,
  title = {Multiphoton emission of x-rays from cooperative resonant nuclei},
  author = {Kong, Xiangjin and Chang, Yue and Zhang, Lida and Yuan, Jianmin and Ma, Yu-Gang},
  journal = {Phys. Rev. Res.},
  volume = {7},
  issue = {1},
  pages = {013030},
  numpages = {7},
  year = {2025},
  month = {Jan},
  publisher = {American Physical Society},
  doi = {10.1103/PhysRevResearch.7.013030},
  url = {https://link.aps.org/doi/10.1103/PhysRevResearch.7.013030}
}

@misc{heeg2016inducingdetectingcollectivepopulation,
      title={Inducing and detecting collective population inversions of {M}\"ossbauer nuclei}, 
      author={K. P. Heeg and C. H. Keitel and J. Evers},
      year={2016},
      eprint={1607.04116},
      archivePrefix={arXiv},
      primaryClass={quant-ph},
}

@book{allen-eberly,
    author = {Allen, L. and Eberly, J.H.} ,
    title = {Optical resonance and two-level atoms},
    publisher = {Dover books on physics and chemistry},
    year = {1987},
    ISBN = {0-486-65533-4}
}

@article{Shvydko2023,
  title = {Resonant X-ray excitation of the nuclear clock isomer ${}^{45}${S}c},
  volume = {622},
  ISSN = {1476-4687},
  url = {http://dx.doi.org/10.1038/s41586-023-06491-w},
  DOI = {10.1038/s41586-023-06491-w},
  number = {7983},
  journal = {Nature},
  publisher = {Springer Science and Business Media LLC},
  author = {Shvyd’ko,  Yuri and R\"{o}hlsberger,  Ralf and Kocharovskaya,  Olga and Evers,  J\"{o}rg and Geloni,  Gianluca Aldo and Liu,  Peifan and Shu,  Deming and Miceli,  Antonino and Stone,  Brandon and Hippler,  Willi and Marx-Glowna,  Berit and Uschmann,  Ingo and Loetzsch,  Robert and Leupold,  Olaf and Wille,  Hans-Christian and Sergeev,  Ilya and Gerharz,  Miriam and Zhang,  Xiwen and Grech,  Christian and Guetg,  Marc and Kocharyan,  Vitali and Kujala,  Naresh and Liu,  Shan and Qin,  Weilun and Zozulya,  Alexey and Hallmann,  J\"{o}rg and Boesenberg,  Ulrike and Jo,  Wonhyuk and M\"{o}ller,  Johannes and Rodriguez-Fernandez,  Angel and Youssef,  Mohamed and Madsen,  Anders and Kolodziej,  Tomasz},
  year = {2023},
  month = sep,
  pages = {471–475}
}

@misc{scandium_new,
      title={Probing the Linewidth of the 12.4-keV Solid-State $^{45}${S}c Isomeric Resonance}, 
      author={Peifan Liu and Miriam Gerharz and Berit Marx-Glowna and Willi Hippler and Jan-Etienne Pudell and Alexey Zozulya and Brandon Stone and Deming Shu and Robert Loetzsch and Sakshath Sadashivaiah and Lars Bocklage and Christina Boemer and Shan Liu and Vitaly Kocharyan and Dietrich Krebs and Tianyun Long and Weilun Qin and Matthias Scholz and Kai Schlage and Ilya Sergeev and Hans-Christian Wille and Ulrike Boesenberg and Gianluca Aldo Geloni and J\"org Hallmann and Wonhyuk Jo and Naresh Kujala and Anders Madsen and Angel Rodriguez-Fernandez and Rustam Rysov and Kelin Tasca and Tomasz Kolodziej and Xiwen Zhang and Markus Ilchen and Niclas Wieland and Günter Huber and James H. Edgar and J\"org Evers and Olga Kocharovskaya and Ralf R\"ohlsberger and Yuri Shvyd'ko},
      year={2025},
      eprint={2508.17538},
      archivePrefix={arXiv},
      primaryClass={quant-ph}, 
}

@misc{single-shot,
   title={Single-shot sorting of {M}\"ossbauer time-domain data at X-ray free electron lasers}, 
   author={Miriam Gerharz and Willi Hippler and Berit Marx-Glowna and Sakshath Sadashivaiah and Kai S. Schulze and Ingo Uschmann and Robert L\"otzsch and Kai Schlage and Sven Velten and Dominik Lentrodt and Lukas Wolff and Olaf Leupold and Ilya Sergeev and Hans-Christian Wille and Cornelius Strohm and Marc Guetg and Shan Liu and Gianluca Aldo Geloni and Ulrike Boesenberg and J\"org Hallmann and Alexey Zozulya and Jan-Etienne Pudell and Angel Rodriguez-Fernandez and Mohamed Youssef and Anders Madsen and Lars Bocklage and Gerhard G. Paulus and Christoph H. Keitel and Thomas Pfeifer and Ralf R\"ohlsberger and J\"org Evers},   
   year={2025},
   eprint={2509.15833},
   archivePrefix={arXiv},
   primaryClass={quant-ph},
}

@article{chumakov2018superradiance,
  title={Superradiance of an ensemble of nuclei excited by a free electron laser},
  author={Chumakov, Aleksandr I and Baron, Alfred QR and Sergueev, Ilya and Strohm, Cornelius and Leupold, Olaf and Shvyd’ko, Yuri and Smirnov, Gennadi V and R{\"u}ffer, Rudolf and Inubushi, Yuichi and Yabashi, Makina and Tono,  Kensuke and Kudo,  Togo and Ishikawa,  Tetsuya},
  journal={Nature physics},
  volume={14},
  number={3},
  pages={261--264},
  year={2018},
  publisher={Nature Publishing Group UK London},
  ISSN = {1745-2481},
  url = {http://dx.doi.org/10.1038/s41567-017-0001-z},
  DOI = {10.1038/s41567-017-0001-z},
}

@article{lentrodt2024towards,
 title = {Toward Nonlinear Optics with {M}\"ossbauer Nuclei Using X-Ray Cavities},
  author = {Lentrodt, Dominik and Keitel, Christoph H. and Evers, J\"org},
  journal = {Phys. Rev. Lett.},
  volume = {135},
  issue = {3},
  pages = {033801},
  numpages = {8},
  year = {2025},
  month = {7},
  publisher = {American Physical Society},
  doi = {10.1103/1lcz-r8d3},
  url = {https://link.aps.org/doi/10.1103/1lcz-r8d3}
}

@misc{adams_scientific_2019,
  title = {Scientific {Opportunities} with an {X}-ray {Free}-{Electron} {Laser} {Oscillator}},
  author = {Adams, Bernhard and Aeppli, Gabriel and Allison, Thomas and Baron, Alfred Q. R. and Bucksbaum, Phillip and Chumakov, Aleksandr I. and Corder, Christopher and Cramer, Stephen P. and DeBeer, Serena and Ding, Yuntao and Evers, J\"org and Frisch, Josef and Fuchs, Matthias and Grübel, Gerhard and Hastings, Jerome B. and Heyl, Christoph M. and Holberg, Leo and Huang, Zhirong and Ishikawa, Tetsuya and Kaldun, Andreas and Kim, Kwang-Je and Kolodziej, Tomasz and Krzywinski, Jacek and Li, Zheng and Liao, Wen-Te and Lindberg, Ryan and Madsen, Anders and Maxwell, Timothy and Monaco, Giulio and Nelson, Keith and Palffy, Adriana and Porat, Gil and Qin, Weilun and Raubenheimer, Tor and Reis, David A. and R\"ohlsberger, Ralf and Santra, Robin and Schoenlein, Robert and Schünemann, Volker and Shpyrko, Oleg and Shvyd'ko, Yuri and Shwartz, Sharon and Singer, Andrej and Sinha, Sunil K. and Sutton, Mark and Tamasaku, Kenji and Wille, Hans-Christian and Yabashi, Makina and Ye, Jun and Zhu, Diling},
  year = {2019},
    url={https://arxiv.org/abs/1903.09317},
   eprint={1903.09317},
   archivePrefix={arXiv},
   primaryClass={physics},
}

@article{Rauer2025,
  title = {Lasing of a cavity-based X-ray source},
  volume = {650},
  ISSN = {1476-4687},
  url = {http://dx.doi.org/10.1038/s41586-025-10025-x},
  DOI = {10.1038/s41586-025-10025-x},
  number = {8100},
  journal = {Nature},
  publisher = {Springer Science and Business Media LLC},
  author = {Rauer,  Patrick and Bahns,  Immo and Friedrich,  Bertram and Casalbuoni,  Sara and Di Felice,  Massimiliano and Dommach,  Martin and Freijo Martin,  Idoia and Freund,  Wolfgang and Gr\"{u}nert,  Jan and Guetg,  Marc and Karpics,  Ivars and Karabekyan,  Suren and Koch,  Andreas and Kujala,  Naresh and La Civita,  Daniele and Liu,  Jia and Maltezopoulos,  Theophilos and Makita,  Mikako and Mayet,  Frank and M\"{u}ller,  Lukas and Rio,  Benoit and Samoylova,  Liubov and Schmidtchen,  Silja and Scholz,  Matthias and Silenzi,  Alessandro and Strauch,  Vivienne and Thoden,  Daniel and Wohlenberg,  Torsten and Vannoni,  Maurizio and Yang,  Fan and Decking,  Winfried and Rossbach,  Joerg and Sinn,  Harald},
  year = {2026},
  month = Jan,
  pages = {93–96}
}

@article{Kagan1979,
  title = {On excitation of isomeric nuclear states in a crystal by synchrotron radiation},
  volume = {12},
  ISSN = {0022-3719},
  url = {http://dx.doi.org/10.1088/0022-3719/12/3/027},
  DOI = {10.1088/0022-3719/12/3/027},
  number = {3},
  journal = {Journal of Physics C: Solid State Physics},
  publisher = {IOP Publishing},
  author = {Kagan,  Yu and Afanas’ev,  A M and Kohn,  V G},
  year = {1979},
  month = feb,
  pages = {615–631}
}

@article{Heeg2013,
  title = {X-ray quantum optics with {M}\"ossbauer nuclei embedded in thin-film cavities},
  author = {Heeg, Kilian P. and Evers, J\"org},
  journal = {Phys. Rev. A},
  volume = {88},
  issue = {4},
  pages = {043828},
  numpages = {12},
  year = {2013},
  month = {10},
  publisher = {American Physical Society},
  doi = {10.1103/PhysRevA.88.043828},
  url = {https://link.aps.org/doi/10.1103/PhysRevA.88.043828}
}

@article{
doi:10.1126/sciadv.adn9825,
author = {Sven Velten  and Lars Bocklage  and Xiwen Zhang  and Kai Schlage  and Anjali Panchwanee  and Sakshath Sadashivaiah  and Ilya Sergeev  and Olaf Leupold  and Aleksandr I. Chumakov  and Olga Kocharovskaya  and Ralf R\"ohlsberger },
title = {Nuclear quantum memory for hard x-ray photon wave packets},
journal = {Science Advances},
volume = {10},
number = {26},
pages = {eadn9825},
year = {2024},
doi = {10.1126/sciadv.adn9825},
URL = {https://www.science.org/doi/abs/10.1126/sciadv.adn9825},
}

@article{PhysRevLett.133.193401,
  title = {Spectral Flux Enhancement of X Rays for Addressing Ultranarrow Nuclear Transitions},
  author = {Kuznetsova, Elena and Zhang, Xiwen and Shvyd'ko, Yuri and Scully, Marlan O. and Kocharovskaya, Olga},
  journal = {Phys. Rev. Lett.},
  volume = {133},
  issue = {19},
  pages = {193401},
  numpages = {6},
  year = {2024},
  month = {Nov},
  publisher = {American Physical Society},
  doi = {10.1103/PhysRevLett.133.193401},
  url = {https://link.aps.org/doi/10.1103/PhysRevLett.133.193401}
}

@article{PhysRevLett.123.250504,
  title = {Nuclear Quantum Memory and Time Sequencing of a Single $\ensuremath{\gamma}$ Photon},
  author = {Zhang, Xiwen and Liao, Wen-Te and Kalachev, Alexey and Shakhmuratov, Rustem and Scully, Marlan and Kocharovskaya, Olga},
  journal = {Phys. Rev. Lett.},
  volume = {123},
  issue = {25},
  pages = {250504},
  numpages = {5},
  year = {2019},
  month = {12},
  publisher = {American Physical Society},
  doi = {10.1103/PhysRevLett.123.250504},
  url = {https://link.aps.org/doi/10.1103/PhysRevLett.123.250504}
}

@article{Rohlsberger2012,
  title = {Electromagnetically induced transparency with resonant nuclei in a cavity},
  volume = {482},
  ISSN = {1476-4687},
  url = {http://dx.doi.org/10.1038/nature10741},
  DOI = {10.1038/nature10741},
  number = {7384},
  journal = {Nature},
  publisher = {Springer Science and Business Media LLC},
  author = {R\"{o}hlsberger,  Ralf and Wille,  Hans-Christian and Schlage,  Kai and Sahoo,  Balaram},
  year = {2012},
  month = feb,
  pages = {199–203}
}

@article{Rohlsberger2010,
author = {Ralf R\"ohlsberger  and Kai Schlage  and Balaram Sahoo  and Sebastien Couet  and Rudolf Rüffer },
title = {Collective Lamb Shift in Single-Photon Superradiance},
journal = {Science},
volume = {328},
number = {5983},
pages = {1248-1251},
year = {2010},
doi = {10.1126/science.1187770},
URL = {https://www.science.org/doi/abs/10.1126/science.1187770}}

@article{Heeg2021,
  title = {Coherent X-ray-optical control of nuclear excitons},
  volume = {590},
  ISSN = {1476-4687},
  url = {http://dx.doi.org/10.1038/s41586-021-03276-x},
  DOI = {10.1038/s41586-021-03276-x},
  number = {7846},
  journal = {Nature},
  publisher = {Springer Science and Business Media LLC},
  author = {Heeg,  Kilian P. and Kaldun,  Andreas and Strohm,  Cornelius and Ott,  Christian and Subramanian,  Rajagopalan and Lentrodt,  Dominik and Haber,  Johann and Wille,  Hans-Christian and Goerttler,  Stephan and R\"{u}ffer,  Rudolf and Keitel,  Christoph H. and R\"{o}hlsberger,  Ralf and Pfeifer,  Thomas and Evers,  J\"{o}rg},
  year = {2021},
  month = feb,
  pages = {401–404}
}

@article{Haber2017,
  title = {Rabi oscillations of X-ray radiation between two nuclear ensembles},
  volume = {11},
  ISSN = {1749-4893},
  url = {http://dx.doi.org/10.1038/s41566-017-0013-3},
  DOI = {10.1038/s41566-017-0013-3},
  number = {11},
  journal = {Nature Photonics},
  publisher = {Springer Science and Business Media LLC},
  author = {Haber,  Johann and Kong,  Xiangjin and Strohm,  Cornelius and Willing,  Svenja and Gollwitzer,  Jakob and Bocklage,  Lars and R\"{u}ffer,  Rudolf and Pálffy,  Adriana and R\"{o}hlsberger,  Ralf},
  year = {2017},
  month = oct,
  pages = {720–725}
}

@article{PhysRevLett.111.073601,
  title = {Vacuum-Assisted Generation and Control of Atomic Coherences at X-Ray Energies},
  author = {Heeg, Kilian P. and Wille, Hans-Christian and Schlage, Kai and Guryeva, Tatyana and Schumacher, Daniel and Uschmann, Ingo and Schulze, Kai S. and Marx, Berit and K\"ampfer, Tino and Paulus, Gerhard G. and R\"ohlsberger, Ralf and Evers, J\"org},
  journal = {Phys. Rev. Lett.},
  volume = {111},
  issue = {7},
  pages = {073601},
  numpages = {5},
  year = {2013},
  month = {8},
  publisher = {American Physical Society},
  doi = {10.1103/PhysRevLett.111.073601},
  url = {https://link.aps.org/doi/10.1103/PhysRevLett.111.073601}
}

@article{PhysRevLett.114.207401,
  title = {Interferometric phase detection at x-ray energies via {Fano} resonance control},
  author = {Heeg, K. P. and Ott, C. and Schumacher, D. and Wille, H.-C. and R\"ohlsberger, R. and Pfeifer, T. and Evers, J.},
  journal = {Phys. Rev. Lett.},
  volume = {114},
  issue = {20},
  pages = {207401},
  numpages = {5},
  year = {2015},
  month = {5},
  publisher = {American Physical Society},
  doi = {10.1103/PhysRevLett.114.207401},
  url = {https://link.aps.org/doi/10.1103/PhysRevLett.114.207401}
}

@article{PhysRevLett.114.203601,
  title = {Tunable Subluminal Propagation of Narrow-band X-Ray Pulses},
  author = {Heeg, Kilian P. and Haber, Johann and Schumacher, Daniel and Bocklage, Lars and Wille, Hans-Christian and Schulze, Kai S. and Loetzsch, Robert and Uschmann, Ingo and Paulus, Gerhard G. and R\"uffer, Rudolf and R\"ohlsberger, Ralf and Evers, J\"org},
  journal = {Phys. Rev. Lett.},
  volume = {114},
  issue = {20},
  pages = {203601},
  numpages = {5},
  year = {2015},
  month = {5},
  publisher = {American Physical Society},
  doi = {10.1103/PhysRevLett.114.203601},
  url = {https://link.aps.org/doi/10.1103/PhysRevLett.114.203601}
}

@article{Haber2016,
  title = {Collective strong coupling of X-rays and nuclei in a nuclear optical lattice},
  volume = {10},
  ISSN = {1749-4893},
  url = {http://dx.doi.org/10.1038/nphoton.2016.77},
  DOI = {10.1038/nphoton.2016.77},
  number = {7},
  journal = {Nature Photonics},
  publisher = {Springer Science and Business Media LLC},
  author = {Haber,  Johann and Schulze,  Kai S. and Schlage,  Kai and Loetzsch,  Robert and Bocklage,  Lars and Gurieva,  Tatiana and Bernhardt,  Hendrik and Wille,  Hans-Christian and R\"{u}ffer,  Rudolf and Uschmann,  Ingo and Paulus,  Gerhard G. and R\"{o}hlsberger,  Ralf},
  year = {2016},
  month = may,
  pages = {445–449}
}

@article{PhysRevLett.66.2037,
  title = {Gamma echo},
  author = {Helist\"o, P. and Tittonen, I. and Lippmaa, M. and Katila, T.},
  journal = {Phys. Rev. Lett.},
  volume = {66},
  issue = {15},
  pages = {2037--2040},
  numpages = {0},
  year = {1991},
  month = {4},
  publisher = {American Physical Society},
  doi = {10.1103/PhysRevLett.66.2037},
  url = {https://link.aps.org/doi/10.1103/PhysRevLett.66.2037}
}

@incollection{yoshida_quantum_2021,
	address = {Singapore},
	title = {Quantum {Optical} {Phenomena} in {Nuclear} {Resonant} {Scattering}},
	volume = {137},
	booktitle = {{Modern} {M{\"o}ssbauer} {Spectroscopy}},
	publisher = {Springer Singapore},
	author = {R{\"o}hlsberger, Ralf and Evers, J{\"o}rg},
	editor = {Yoshida, Yutaka and Langouche, Guido},
	year = {2021},
	note = {Series Title: Topics in Applied Physics},
	pages = {105--171},
    url="https://doi.org/10.1007/978-981-15-9422-9_3"
}

@misc{adaptiveOptics,
      title={Dark-fringe interferometer with dynamic phase control for {M}\"ossbauer science}, 
      author={Miriam Gerharz and Dominik Lentrodt and Lars Bocklage and Kai Schulze and Christian Ott and René Steinbrügge and Olaf Leupold and Ilya Sergeev and Gerhard G. Paulus and Christoph H. Keitel and Ralf Röhlsberger and Thomas Pfeifer and Jörg Evers},
      year={2025},
      eprint={2509.24658},
      archivePrefix={arXiv},
      primaryClass={quant-ph},
}

@incollection{moessbauer_story_dreams,
  author      = {Shenoy, G.},
  title       = {Dreams with synchrotron radiation},
  editor      = {Kalvius, G. and Kienle, P.},
  booktitle   = {The Rudolf {M}\"ossbauer Story: His Scientific Work and Its Impact on Science and History},
  publisher   = {Springer},
  address     = { Berlin, Heidelberg},
  isbn        = {9783642179525},
  year        = 2012
}

@article{PhysRevLett.100.244802,
  title = {A Proposal for an X-Ray Free-Electron Laser Oscillator with an Energy-Recovery Linac},
  author = {Kim, Kwang-Je and Shvyd'ko, Yuri and Reiche, Sven},
  journal = {Phys. Rev. Lett.},
  volume = {100},
  issue = {24},
  pages = {244802},
  numpages = {4},
  year = {2008},
  month = {6},
  publisher = {American Physical Society},
  doi = {10.1103/PhysRevLett.100.244802},
  url = {https://link.aps.org/doi/10.1103/PhysRevLett.100.244802}
}

@article{Margraf2023,
  title = {Low-loss stable storage of 1.2\,Å X-ray pulses in a 14 m Bragg cavity},
  volume = {17},
  ISSN = {1749-4893},
  url = {http://dx.doi.org/10.1038/s41566-023-01267-0},
  DOI = {10.1038/s41566-023-01267-0},
  number = {10},
  journal = {Nature Photonics},
  publisher = {Springer Science and Business Media LLC},
  author = {Margraf,  Rachel and Robles,  River and Halavanau,  Alex and Kryzywinski,  Jacek and Li,  Kenan and MacArthur,  James and Osaka,  Taito and Sakdinawat,  Anne and Sato,  Takahiro and Sun,  Yanwen and Tamasaku,  Kenji and Huang,  Zhirong and Marcus,  Gabriel and Zhu,  Diling},
  year = {2023},
  month = aug,
  pages = {878–882}
}

@article{10.1063/1.5138937,
    author = {Sánchez-Barquilla, M. and Silva, R. E. F. and Feist, J.},
    title = {Cumulant expansion for the treatment of light–matter interactions in arbitrary material structures},
    journal = {The Journal of Chemical Physics},
    volume = {152},
    number = {3},
    pages = {034108},
    year = {2020},
    month = {01},
    issn = {0021-9606},
    doi = {10.1063/1.5138937},
    url = {https://doi.org/10.1063/1.5138937},
}

@article{PhysRevResearch.5.013091,
  title = {Characterizing superradiant dynamics in atomic arrays via a cumulant expansion approach},
  author = {Rubies-Bigorda, Oriol and Ostermann, Stefan and Yelin, Susanne F.},
  journal = {Phys. Rev. Res.},
  volume = {5},
  issue = {1},
  pages = {013091},
  numpages = {12},
  year = {2023},
  month = {Feb},
  publisher = {American Physical Society},
  doi = {10.1103/PhysRevResearch.5.013091},
  url = {https://link.aps.org/doi/10.1103/PhysRevResearch.5.013091}
}

@article{Kirton_2018,
doi = {10.1088/1367-2630/aaa11d},
url = {https://dx.doi.org/10.1088/1367-2630/aaa11d},
year = {2018},
month = {jan},
publisher = {IOP Publishing},
volume = {20},
number = {1},
pages = {015009},
author = {Kirton, Peter and Keeling, Jonathan},
title = {Superradiant and lasing states in driven-dissipative Dicke models},
journal = {New Journal of Physics}
}

@article{Krmer2015,
  title = {Generalized mean-field approach to simulate the dynamics of large open spin ensembles with long range interactions},
  volume = {69},
  ISSN = {1434-6079},
  url = {http://dx.doi.org/10.1140/epjd/e2015-60266-5},
  DOI = {10.1140/epjd/e2015-60266-5},
  number = {12},
  journal = {The European Physical Journal D},
  publisher = {Springer Science and Business Media LLC},
  author = {Kr\"{a}mer,  Sebastian and Ritsch,  Helmut},
  year = {2015},
  month = dec 
}

@article{Lentrodt2020b,
  title = {Ab initio quantum models for thin-film x-ray cavity QED},
  author = {Lentrodt, Dominik and Heeg, Kilian P. and Keitel, Christoph H. and Evers, J\"org},
  journal = {Phys. Rev. Res.},
  volume = {2},
  issue = {2},
  pages = {023396},
  numpages = {38},
  year = {2020},
  month = {6},
  publisher = {American Physical Society},
  doi = {10.1103/PhysRevResearch.2.023396},
  url = {https://link.aps.org/doi/10.1103/PhysRevResearch.2.023396}
}

@article{PhysRevA.102.033710,
  title = {Green's-function formalism for resonant interaction of x rays with nuclei in structured media},
  author = {Kong, Xiangjin and Chang, Darrick E. and P\'alffy, Adriana},
  journal = {Phys. Rev. A},
  volume = {102},
  issue = {3},
  pages = {033710},
  numpages = {10},
  year = {2020},
  month = {Sep},
  publisher = {American Physical Society},
  doi = {10.1103/PhysRevA.102.033710},
  url = {https://link.aps.org/doi/10.1103/PhysRevA.102.033710}
}

@article{PhysRevLett.77.3232,
  title = {Storage of Nuclear Excitation Energy through Magnetic Switching},
  author = {Shvyd'ko, Yu. V. and Hertrich, T. and van B\"urck, U. and Gerdau, E. and Leupold, O. and Metge, J. and R\"uter, H. D. and Schwendy, S. and Smirnov, G. V. and Potzel, W. and Schindelmann, P.},
  journal = {Phys. Rev. Lett.},
  volume = {77},
  issue = {15},
  pages = {3232--3235},
  numpages = {0},
  year = {1996},
  month = {10},
  publisher = {American Physical Society},
  doi = {10.1103/PhysRevLett.77.3232},
  url = {https://link.aps.org/doi/10.1103/PhysRevLett.77.3232}
}

@article{PhysRevA.65.023804,
  title = {Radiative decoupling and coupling of nuclear oscillators by stepwise Doppler-energy shifts},
  author = {Schindelmann, P. and van B\"urck, U. and Potzel, W. and Smirnov, G. V. and Popov, S. L. and Gerdau, E. and Shvyd'ko, Yu. V. and J\"aschke, J. and R\"uter, H. D. and Chumakov, A. I. and R\"uffer, R.},
  journal = {Phys. Rev. A},
  volume = {65},
  issue = {2},
  pages = {023804},
  numpages = {17},
  year = {2002},
  month = {1},
  publisher = {American Physical Society},
  doi = {10.1103/PhysRevA.65.023804},
  url = {https://link.aps.org/doi/10.1103/PhysRevA.65.023804}
}

@article{Vagizov2014,
  title = {Coherent control of the waveforms of recoilless $\gamma$-ray photons},
  volume = {508},
  ISSN = {1476-4687},
  url = {http://dx.doi.org/10.1038/nature13018},
  DOI = {10.1038/nature13018},
  number = {7494},
  journal = {Nature},
  publisher = {Springer Science and Business Media LLC},
  author = {Vagizov,  Farit and Antonov,  Vladimir and Radeonychev,  Y. V. and Shakhmuratov,  R. N. and Kocharovskaya,  Olga},
  year = {2014},
  month = mar,
  pages = {80–83}
}

@Article{Vagizov2013,
author={Vagizov, F. G.
and Sadykov, E. K.
and Kocharovskaya, O. A.},
title={Modulation of {M}{\"o}ssbauer radiation by pulsed laser excitation},
journal={JETP Letters},
year={2013},
month={2},
day={01},
volume={96},
number={12},
pages={812-816},
issn={1090-6487},
doi={10.1134/S0021364012240137},
url={https://doi.org/10.1134/S0021364012240137}
}

@Article{Sakshath2017,
author="Sakshath, S.
and Jenni, K.
and Scherthan, L.
and W{\"u}rtz, P.
and Herlitschke, M.
and Sergeev, I.
and Strohm, C.
and Wille, H.-C.
and R{\"o}hlsberger, R.
and Wolny, J. A.
and Sch{\"u}nemann, V.",
title="Optical pump - nuclear resonance probe experiments on spin crossover complexes",
journal="Hyperfine Interactions",
year="2017",
month="Nov",
day="13",
volume="238",
number="1",
pages="89",
issn="1572-9540",
doi="10.1007/s10751-017-1461-3",
nourl ="https://doi.org/10.1007/s10751-017-1461-3"
}

@article{PhysRevA.95.033818,
  title = {Atom-light interactions in quasi-one-dimensional nanostructures: A Green's-function perspective},
  author = {Asenjo-Garcia, A. and Hood, J. D. and Chang, D. E. and Kimble, H. J.},
  journal = {Phys. Rev. A},
  volume = {95},
  issue = {3},
  pages = {033818},
  numpages = {16},
  year = {2017},
  month = {Mar},
  publisher = {American Physical Society},
  doi = {10.1103/PhysRevA.95.033818},
  url = {https://link.aps.org/doi/10.1103/PhysRevA.95.033818}
}

@Book{ralf,
	author="R{\"o}hlsberger, Ralf",
	title="Coherent Elastic Nuclear Resonant Scattering",
	bookTitle="Nuclear Condensed Matter Physics with Synchrotron Radiation: Basic Principles, Methodology and Applications",
	year="2004",
	publisher="Springer",
	address="Berlin, Heidelberg"
}

@article{lentrodt2024excitationnarrowxraytransitions,
      title = {Excitation of narrow x-ray transitions in thin-film cavities by focused pulses},
  author = {Lentrodt, Dominik and Keitel, Christoph H. and Evers, J\"org},
  journal = {Phys. Rev. A},
  volume = {112},
  issue = {1},
  pages = {013711},
  numpages = {17},
  year = {2025},
  month = {7},
  publisher = {American Physical Society},
  doi = {10.1103/PhysRevA.112.013711},
  url = {https://link.aps.org/doi/10.1103/PhysRevA.112.013711}
}

@article{emma2010first,
  title={First lasing and operation of an {\aa}ngstrom-wavelength free-electron laser},
  author = {Emma,  P. and Akre,  R. and Arthur,  J. and Bionta,  R. and Bostedt,  C. and Bozek,  J. and Brachmann,  A. and Bucksbaum,  P. and Coffee,  R. and Decker,  F.-J. and Ding,  Y. and Dowell,  D. and Edstrom,  S. and Fisher,  A. and Frisch,  J. and Gilevich,  S. and Hastings,  J. and Hays,  G. and Hering,  Ph. and Huang,  Z. and Iverson,  R. and Loos,  H. and Messerschmidt,  M. and Miahnahri,  A. and Moeller,  S. and Nuhn,  H.-D. and Pile,  G. and Ratner,  D. and Rzepiela,  J. and Schultz,  D. and Smith,  T. and Stefan,  P. and Tompkins,  H. and Turner,  J. and Welch,  J. and White,  W. and Wu,  J. and Yocky,  G. and Galayda,  J.},
  journal={Nature Photonics},
  volume={4},
  number={9},
  pages={641--647},
  year={2010},
  publisher={Nature Publishing Group UK London},
  url = {http://dx.doi.org/10.1038/nphoton.2010.176},
}

@article{amann2012demonstration,
  title={Demonstration of self-seeding in a hard-X-ray free-electron laser},
  author={Amann, J and Berg, W and Blank, V and Decker, F-J and Ding, Y and Emma, P and Feng, Y and Frisch, J and Fritz, D and Hastings, J and others},
  journal={Nature photonics},
  volume={6},
  number={10},
  pages={693--698},
  year={2012},
  publisher={Nature Publishing Group UK London},
  url = {http://dx.doi.org/10.1038/nphoton.2012.180},
}

@article{ishikawa2012compact,
  title={A compact X-ray free-electron laser emitting in the sub-{\aa}ngstr{\"o}m region},
  author = {Ishikawa,  Tetsuya and Aoyagi,  Hideki and Asaka,  Takao and Asano,  Yoshihiro and Azumi,  Noriyoshi and Bizen,  Teruhiko and Ego,  Hiroyasu and Fukami,  Kenji and Fukui,  Toru and Furukawa,  Yukito and Goto,  Shunji and Hanaki,  Hirofumi and Hara,  Toru and Hasegawa,  Teruaki and Hatsui,  Takaki and Higashiya,  Atsushi and Hirono,  Toko and Hosoda,  Naoyasu and Ishii,  Miho and Inagaki,  Takahiro and Inubushi,  Yuichi and Itoga,  Toshiro and Joti,  Yasumasa and Kago,  Masahiro and Kameshima,  Takashi and Kimura,  Hiroaki and Kirihara,  Yoichi and Kiyomichi,  Akio and Kobayashi,  Toshiaki and Kondo,  Chikara and Kudo,  Togo and Maesaka,  Hirokazu and Maréchal,  Xavier M. and Masuda,  Takemasa and Matsubara,  Shinichi and Matsumoto,  Takahiro and Matsushita,  Tomohiro and Matsui,  Sakuo and Nagasono,  Mitsuru and Nariyama,  Nobuteru and Ohashi,  Haruhiko and Ohata,  Toru and Ohshima,  Takashi and Ono,  Shun and Otake,  Yuji and Saji,  Choji and Sakurai,  Tatsuyuki and Sato,  Takahiro and Sawada,  Kei and Seike,  Takamitsu and Shirasawa,  Katsutoshi and Sugimoto,  Takashi and Suzuki,  Shinsuke and Takahashi,  Sunao and Takebe,  Hideki and Takeshita,  Kunikazu and Tamasaku,  Kenji and Tanaka,  Hitoshi and Tanaka,  Ryotaro and Tanaka,  Takashi and Togashi,  Tadashi and Togawa,  Kazuaki and Tokuhisa,  Atsushi and Tomizawa,  Hiromitsu and Tono,  Kensuke and Wu,  Shukui and Yabashi,  Makina and Yamaga,  Mitsuhiro and Yamashita,  Akihiro and Yanagida,  Kenichi and Zhang,  Chao and Shintake,  Tsumoru and Kitamura,  Hideo and Kumagai,  Noritaka},
  journal={Nature Photonics},
  volume={6},
  number={8},
  pages={540--544},
  year={2012},
  publisher={Nature Publishing Group UK London},
  url = {http://dx.doi.org/10.1038/nphoton.2012.141},
}

@article{inoue2019generation,
  title={Generation of narrow-band X-ray free-electron laser via reflection self-seeding},
  author = {Inoue,  Ichiro and Osaka,  Taito and Hara,  Toru and Tanaka,  Takashi and Inagaki,  Takahiro and Fukui,  Toru and Goto,  Shunji and Inubushi,  Yuichi and Kimura,  Hiroaki and Kinjo,  Ryota and Ohashi,  Haruhiko and Togawa,  Kazuaki and Tono,  Kensuke and Yamaga,  Mitsuhiro and Tanaka,  Hitoshi and Ishikawa,  Tetsuya and Yabashi,  Makina},
  journal={Nature photonics},
  volume={13},
  number={5},
  pages={319--322},
  year={2019},
  publisher={Nature Publishing Group UK London},
  url = {http://dx.doi.org/10.1038/s41566-019-0365-y},
}

@Article{Hannon1999,
	author={Hannon, J. P.
	and Trammell, G. T.},
	title={Coherent $\ensuremath{\gamma}$-ray optics},
	journal={Hyperfine Interactions},
	year={1999},
	volume={123},
	number={1},
	pages={127-274},
	issn={1572-9540},
	doi={10.1023/A:1017011621007},
	url={https://doi.org/10.1023/A:1017011621007}
}

@BOOK{Ficek_Swain,
    title = {Quantum interference and coherence: theory and experiments},
    publisher = {Springer},
    year = {2005},
    address= {Heidelberg},
    author = {Ficek, Z. and Swain, S.},
    series = {Springer series in optical sciences}
}

@INCOLLECTION{Kiffner_Vacuum_Processes,
  author = {Martin Kiffner and M. Macovei and J. Evers and C. H. Keitel},
  title = {Vacuum-induced processes in multi-level atoms},
  booktitle = {Progress in Optics},
  publisher = {Elsevier Science},
  year = {2010},
  volume = {55},
  pages = {85-197},
  address = {Burlington},
  url = {https://doi.org/10.1016/B978-0-444-53705-8.00003-5}
}

@article{PhysRevA.104.023702,
  title = {Beyond lowest order mean-field theory for light interacting with atom arrays},
  author = {Robicheaux, F. and Suresh, Deepak A.},
  journal = {Phys. Rev. A},
  volume = {104},
  issue = {2},
  pages = {023702},
  numpages = {12},
  year = {2021},
  month = {Aug},
  publisher = {American Physical Society},
  doi = {10.1103/PhysRevA.104.023702},
  url = {https://link.aps.org/doi/10.1103/PhysRevA.104.023702}
}

@article{PhysRevA.78.022102,
  title = {Cluster-expansion representation in quantum optics},
  author = {Kira, M. and Koch, S. W.},
  journal = {Phys. Rev. A},
  volume = {78},
  issue = {2},
  pages = {022102},
  numpages = {26},
  year = {2008},
  month = {Aug},
  publisher = {American Physical Society},
  doi = {10.1103/PhysRevA.78.022102},
  url = {https://link.aps.org/doi/10.1103/PhysRevA.78.022102}
}

@article{SCHOLLWOCK201196,
title = {The density-matrix renormalization group in the age of matrix product states},
journal = {Annals of Physics},
volume = {326},
number = {1},
pages = {96-192},
year = {2011},
note = {January 2011 Special Issue},
issn = {0003-4916},
doi = {https://doi.org/10.1016/j.aop.2010.09.012},
url = {https://www.sciencedirect.com/science/article/pii/S0003491610001752},
author = {Ulrich Schollwöck}
}

@article{nam2021high,
  title={High-brightness self-seeded X-ray free-electron laser covering the 3.5 ke{V} to 14.6 ke{V} range},
 author = {Nam,  Inhyuk and Min,  Chang-Ki and Oh,  Bonggi and Kim,  Gyujin and Na,  Donghyun and Suh,  Young Jin and Yang,  Haeryong and Cho,  Myung Hoon and Kim,  Changbum and Kim,  Min-Jae and Shim,  Chi Hyun and Ko,  Jun Ho and Heo,  Hoon and Park,  Jaehyun and Kim,  Jangwoo and Park,  Sehan and Park,  Gisu and Kim,  Seonghan and Chun,  Sae Hwan and Hyun,  HyoJung and Lee,  Jae Hyuk and Kim,  Kyung Sook and Eom,  Intae and Rah,  Seungyu and Shu,  Deming and Kim,  Kwang-Je and Terentyev,  Sergey and Blank,  Vladimir and Shvyd’ko,  Yuri and Lee,  Sang Jae and Kang,  Heung-Sik},
  journal={Nature Photonics},
  volume={15},
  number={6},
  pages={435--441},
  year={2021},
  publisher={Nature Publishing Group UK London},
  url = {https://doi.org/10.1038/s41566-021-00777-z}
}

@book{greenwood2012mossbauer,
  title={M{\"o}ssbauer spectroscopy},
  author={Greenwood, Norman Neill},
  year={2012},
  publisher={Springer Science \& Business Media}
}

@article{barletta2010free,
  title={Free electron lasers: Present status and future challenges},
  author = {W.A. Barletta and J. Bisognano and J.N. Corlett and P. Emma and Z. Huang and K.-J. Kim and R. Lindberg and J.B. Murphy and G.R. Neil and D.C. Nguyen and C. Pellegrini and R.A. Rimmer and F. Sannibale and G. Stupakov and R.P. Walker and A.A. Zholents},
  journal={Nuclear Instruments and Methods in Physics Research Section A: Accelerators, Spectrometers, Detectors and Associated Equipment},
  volume={618},
  number={1-3},
  pages={69--96},
  year={2010},
  publisher={Elsevier},
url = {https://www.sciencedirect.com/science/article/pii/S0168900210005656},
}

@article{decking2020mhz,
  title={A {MH}z-repetition-rate hard X-ray free-electron laser driven by a superconducting linear accelerator},
  author={Decking, Winfried and Abeghyan, S and Abramian, P and Abramsky, A and Aguirre, A and Albrecht, C and Alou, P and Altarelli, M and Altmann, P and Amyan, K and others},
  journal={Nature photonics},
  volume={14},
  number={6},
  pages={391--397},
  year={2020},
  publisher={Nature Publishing Group UK London},
  url = {http://dx.doi.org/10.1038/s41566-020-0607-z},
}

@article{liu2023cascaded,
  title={Cascaded hard X-ray self-seeded free-electron laser at megahertz repetition rate},
  author = {Liu,  Shan and Grech,  Christian and Guetg,  Marc and Karabekyan,  Suren and Kocharyan,  Vitali and Kujala,  Naresh and Lechner,  Christoph and Long,  Tianyun and Mirian,  Najmeh and Qin,  Weilun and Serkez,  Svitozar and Tomin,  Sergey and Yan,  Jiawei and Abeghyan,  Suren and Anton,  Jayson and Blank,  Vladimir and Boesenberg,  Ulrike and Brinker,  Frank and Chen,  Ye and Decking,  Winfried and Dong,  Xiaohao and Kearney,  Steve and La Civita,  Daniele and Madsen,  Anders and Maltezopoulos,  Theophilos and Rodriguez-Fernandez,  Angel and Saldin,  Evgeni and Samoylova,  Liubov and Scholz,  Matthias and Sinn,  Harald and Sleziona,  Vivien and Shu,  Deming and Tanikawa,  Takanori and Terentiev,  Sergey and Trebushinin,  Andrei and Tschentscher,  Thomas and Vannoni,  Maurizio and Wohlenberg,  Torsten and Yakopov,  Mikhail and Geloni,  Gianluca},
  journal={Nature Photonics},
  volume={17},
  number={11},
  pages={984--991},
  year={2023},
  publisher={Nature Publishing Group UK London},
url = {https://doi.org/10.1038/s41566-023-01305-x}
}

@article{Liu04032025,
author = {Shan Liu and Gianluca Geloni and Tianyun Long and Weilun Qin and Vitali Kocharyan and Jiawei Yan and Lu Cao and Naresh Kujala and Marc Guetg and Matthias Scholz and Winfried Decking},
title = {Updates on the Hard X-Ray Self-Seeding at the European XFEL},
journal = {Synchrotron Radiation News},
volume = {38},
number = {2},
pages = {11--16},
year = {2025},
publisher = {Taylor \& Francis},
doi = {10.1080/08940886.2025.2472607}
}

@article{nazeeri2025couplingnucleartransitionsurface,
  title = {Coupling of a Nuclear Transition to a Surface Acoustic Wave},
  author = {Nazeeri, Albert and Brandenstein, Chiara and Jia, Chengjie and Magrini, Lorenzo and Gratta, Giorgio},
  journal = {Phys. Rev. Lett.},
  volume = {136},
  issue = {18},
  pages = {183801},
  numpages = {5},
  year = {2026},
  month = {May},
  publisher = {American Physical Society},
  doi = {10.1103/tc97-98f7},
  url = {https://link.aps.org/doi/10.1103/tc97-98f7}
}

@article{Yamashita2024,
  title = {The correlation of the gamma ray waveform with the vibration phase of the resonant absorber},
  volume = {245},
  ISSN = {3005-0731},
  url = {http://dx.doi.org/10.1007/s10751-024-02214-3},
  DOI = {10.1007/s10751-024-02214-3},
  number = {1},
  journal = {Interactions},
  publisher = {Springer Science and Business Media LLC},
  author = {Yamashita,  Hiroyuki and Kitao,  Shinji and Kobayashi,  Yasuhiro and Seto,  Makoto},
  year = {2024},
  month = nov 
}

@article{10.1063/5.0249167,
    author = {Stejskal, Aleš and Vrba, Vlastimil and Procházka, Vít},
    title = {Toward flexible intensity control of resonantly scattered $\gamma$-rays using multi-frequency vibrating resonant absorber},
    journal = {Applied Physics Letters},
    volume = {126},
    number = {8},
    pages = {084102},
    year = {2025},
    month = {02},
    issn = {0003-6951},
    doi = {10.1063/5.0249167},
    url = {https://doi.org/10.1063/5.0249167},
    }

@article{doi:10.7566/JPSJ.90.084705,
author = {Fujiwara ,Kosuke and Mitsui ,Takaya and Aoyagi ,Yumito and Yoda ,Yoshitaka and Ikeda ,Naoshi},
title = {Quantum Interference of Totally Reflected Mössbauer $\gamma$-Rays from a 57Fe Monolayer Embedded in a Thin Film},
journal = {Journal of the Physical Society of Japan},
volume = {90},
number = {8},
pages = {084705},
year = {2021},
doi = {10.7566/JPSJ.90.084705},
URL = {   https://doi.org/10.7566/JPSJ.90.084705}
}

@article{JBWang_2004,
doi = {10.1088/0957-4484/15/5/014},
url = {https://doi.org/10.1088/0957-4484/15/5/014},
year = {2004},
month = {feb},
publisher = {},
volume = {15},
number = {5},
pages = {485},
author = {J B Wang and X Z Zhou and Q F Liu and D S Xue and F S Li and B Li and H P Kunkel and G Williams},
title = {Magnetic texture in iron nanowire arrays},
journal = {Nanotechnology}
}

@article{PhysRevB.72.024428,
  title = {Dipolar interactions in arrays of iron nanowires studied by M\"ossbauer spectroscopy},
  author = {Zhan, Qing-Feng and Gao, Jian-Hua and Liang, Ya-Qiong and Di, Na-Li and Cheng, Zhao-Hua},
  journal = {Phys. Rev. B},
  volume = {72},
  issue = {2},
  pages = {024428},
  numpages = {5},
  year = {2005},
  month = {Jul},
  publisher = {American Physical Society},
  doi = {10.1103/PhysRevB.72.024428},
  url = {https://link.aps.org/doi/10.1103/PhysRevB.72.024428}
}

@article{10.1063/1.373001,
    author = {Peng, Yong and Zhang, Hao-Li and Pan, Shan-Lin and Li, Hu-Lin},
    title = {Magnetic properties and magnetization reversal of $\alpha$-Fe nanowires deposited in alumina film},
    journal = {Journal of Applied Physics},
    volume = {87},
    number = {10},
    pages = {7405-7408},
    year = {2000},
    month = {05},
    issn = {0021-8979},
    doi = {10.1063/1.373001},
}

@ARTICLE{824421,
  author={Metzger, R.M. and Konovalov, V.V. and Ming Sun and Tao Xu and Zangari, G. and Bin Xu and Benakli, M. and Doyle, W.D.},
  journal={IEEE Transactions on Magnetics}, 
  title={Magnetic nanowires in hexagonally ordered pores of alumina}, 
  year={2000},
  volume={36},
  number={1},
  pages={30-35},
  doi={10.1109/20.824421}
}

@Article{Mohaddes-Ardabili2004,
author={Mohaddes-Ardabili, L.
and Zheng, H.
and Ogale, S. B.
and Hannoyer, B.
and Tian, W.
and Wang, J.
and Lofland, S. E.
and Shinde, S. R.
and Zhao, T.
and Jia, Y.
and Salamanca-Riba, L.
and Schlom, D. G.
and Wuttig, M.
and Ramesh, R.},
title={Self-assembled single-crystal ferromagnetic iron nanowires formed by decomposition},
journal={Nature Materials},
year={2004},
month={Aug},
day={01},
volume={3},
number={8},
pages={533-538},
issn={1476-4660},
doi={10.1038/nmat1162},
url={https://doi.org/10.1038/nmat1162}
}

@Article{Grinter2023,
author={Grinter, David C.
and Shaw, Bobbie-Jean A.
and Pang, Chi L.
and Yim, Chi-Ming
and Muryn, Christopher A.
and Hall, Charlotte A.
and Maccherozzi, Francesco
and Dhesi, Sarnjeet S.
and Suzuki, Masahiko
and Yasue, Tsuneo
and Koshikawa, Takanori
and Thornton, Geoff},
title={Fabrication of Isolated Iron Nanowires},
journal={The Journal of Physical Chemistry Letters},
year={2023},
month={Sep},
day={28},
publisher={American Chemical Society},
volume={14},
number={38},
pages={8507-8512},
doi={10.1021/acs.jpclett.3c02362},
url={https://doi.org/10.1021/acs.jpclett.3c02362}
}

@article{zs9x-9x6f,
  title = {Symmetry-based efficient simulation of higher-order coherences in quantum many-body superradiance},
  author = {Holzinger, Raphael and Rubies-Bigorda, Oriol and Yelin, Susanne F. and Ritsch, Helmut},
  journal = {Phys. Rev. Res.},
  volume = {7},
  issue = {3},
  pages = {033203},
  numpages = {10},
  year = {2025},
  month = {Aug},
  publisher = {American Physical Society},
  doi = {10.1103/zs9x-9x6f},
  url = {https://link.aps.org/doi/10.1103/zs9x-9x6f}
}

@article{Krämer_2016,
doi = {10.1209/0295-5075/114/14003},
url = {https://doi.org/10.1209/0295-5075/114/14003},
year = {2016},
month = {may},
publisher = {EDP Sciences, IOP Publishing and Società Italiana di Fisica},
volume = {114},
number = {1},
pages = {14003},
author = {Krämer, S. and Ostermann, L. and Ritsch, H.},
title = {Optimized geometries for future generation optical lattice clocks},
journal = {Europhysics Letters}
}

@article{Erb2022,
  title = {Real-Time Observation of Temperature-Induced Surface Nanofaceting in M-Plane $\alpha$-Al2O3},
  volume = {14},
  ISSN = {1944-8252},
  url = {http://dx.doi.org/10.1021/acsami.1c22029},
  DOI = {10.1021/acsami.1c22029},
  number = {27},
  journal = {ACS Applied Materials \& Interfaces},
  publisher = {American Chemical Society (ACS)},
  author = {Erb,  Denise J. and Perlich,  Jan and Roth,  Stephan V. and R\"{o}hlsberger,  Ralf and Schlage,  Kai},
  year = {2022},
  month = jun,
  pages = {31373–31384}
}

@article{PhysRevLett.118.237204,
  title = {Nuclear Resonant Surface Diffraction of Synchrotron Radiation},
  author = {Schlage, Kai and Dzemiantsova, Liudmila and Bocklage, Lars and Wille, Hans-Christian and Pues, Matthias and Meier, Guido and R\"ohlsberger, Ralf},
  journal = {Phys. Rev. Lett.},
  volume = {118},
  issue = {23},
  pages = {237204},
  numpages = {6},
  year = {2017},
  month = {Jun},
  publisher = {American Physical Society},
  doi = {10.1103/PhysRevLett.118.237204},
  url = {https://link.aps.org/doi/10.1103/PhysRevLett.118.237204}
}

@article{Ostermann2019,
  title = {Super- and subradiance of clock atoms in multimode optical waveguides},
  volume = {21},
  ISSN = {1367-2630},
  url = {http://dx.doi.org/10.1088/1367-2630/ab05fb},
  DOI = {10.1088/1367-2630/ab05fb},
  number = {2},
  journal = {New Journal of Physics},
  publisher = {IOP Publishing},
  author = {Ostermann,  Laurin and Meignant,  Clément and Genes,  Claudiu and Ritsch,  Helmut},
  year = {2019},
  month = feb,
  pages = {025004}
}

@Article{photonics12100996,
AUTHOR = {Fasser, Martin and Genes, Claudiu and Ritsch, Helmut and Holzinger, Raphael},
TITLE = {Benchmarking the Cumulant Expansion Method Using Dicke Superradiance},
JOURNAL = {Photonics},
VOLUME = {12},
YEAR = {2025},
NUMBER = {10},
ARTICLE-NUMBER = {996},
URL = {https://www.mdpi.com/2304-6732/12/10/996},
ISSN = {2304-6732},
DOI = {10.3390/photonics12100996}
}

@article{plankensteiner2022quantumcumulants,
  doi = {10.22331/q-2022-01-04-617},
  url = {https://doi.org/10.22331/q-2022-01-04-617},
  title = {Quantum{C}umulants.jl: {A} {J}ulia framework for generalized mean-field equations in open quantum systems},
  author = {Plankensteiner, David and Hotter, Christoph and Ritsch, Helmut},
  journal = {{Quantum}},
  issn = {2521-327X},
  publisher = {{Verein zur F{\"{o}}rderung des Open Access Publizierens in den Quantenwissenschaften}},
  volume = {6},
  pages = {617},
  month = jan,
  year = {2022}
}

@article{Lohse2026,
  title = {Interferometric measurement of nuclear resonant phase shift with a nanoscale Young double waveguide},
  ISSN = {1749-4893},
  url = {http://dx.doi.org/10.1038/s41566-026-01892-5},
  DOI = {10.1038/s41566-026-01892-5},
  journal = {Nature Photonics},
  publisher = {Springer Science and Business Media LLC},
  author = {Lohse,  Leon M. and Negi,  Ankita and Osterhoff,  Markus and Meyer,  Paul and Yaroslavtsev,  Sergey and Chumakov,  Aleksandr I. and Bocklage,  Lars and R\"{o}hlsberger,  Ralf and Salditt,  Tim},
  year = {2026},
  month = apr 
}

@article{Henke1993,
  author  = {B. L. Henke and E. M. Gullikson and J. C. Davis},
  title   = {X-ray interactions: photoabsorption, scattering, transmission, and reflection at $E=50-30000$ eV, $Z=1-92$},
  journal = {Atomic Data and Nuclear Data Tables},
  volume  = {54},
  number  = {2},
  pages   = {181--342},
  year    = {1993},
  month   = {July},
  doi     = {10.1006/adnd.1993.1013},
  url     = {https://doi.org/10.1006/adnd.1993.1013}
}

@article{PhysRevA.100.041602,
  title = {Spin squeezing and many-body dipolar dynamics in optical lattice clocks},
  author = {Qu, Chunlei and Rey, Ana M.},
  journal = {Phys. Rev. A},
  volume = {100},
  issue = {4},
  pages = {041602(R)},
  numpages = {7},
  year = {2019},
  month = {Oct},
  publisher = {American Physical Society},
  doi = {10.1103/PhysRevA.100.041602},
  url = {https://link.aps.org/doi/10.1103/PhysRevA.100.041602}
}

@article{PRXQuantum.5.010344,
  title = {Dicke Superradiance in Ordered Arrays of Multilevel Atoms},
  author = {Masson, Stuart J. and Covey, Jacob P. and Will, Sebastian and Asenjo-Garcia, Ana},
  journal = {PRX Quantum},
  volume = {5},
  issue = {1},
  pages = {010344},
  numpages = {19},
  year = {2024},
  month = {Mar},
  publisher = {American Physical Society},
  doi = {10.1103/PRXQuantum.5.010344},
  url = {https://link.aps.org/doi/10.1103/PRXQuantum.5.010344}
}

@article{Zhu_2015,
doi = {10.1088/1367-2630/17/8/083063},
url = {https://doi.org/10.1088/1367-2630/17/8/083063},
year = {2015},
month = {sep},
publisher = {IOP Publishing},
volume = {17},
number = {8},
pages = {083063},
author = {Zhu, B and Schachenmayer, J and Xu, M and Herrera, F and Restrepo, J G and Holland, M J and Rey, A M},
title = {Synchronization of interacting quantum dipoles},
journal = {New Journal of Physics}
}

\end{document}